\documentclass[]{aa}  
\usepackage{graphicx}
\usepackage{natbib}
\usepackage{txfonts}
\usepackage[colorlinks = true,linkcolor=blue,citecolor=blue,urlcolor = blue]{hyperref} 
\newcommand{\stars}{Cyg~OB2~\#8A}

\begin{document}

\title{Search for non-thermal X-ray emission in the colliding wind binary \stars{}\thanks{Based on data collected with NASA missions {\it NuSTAR} and {\it Swift}, and the ESA observatories {\it XMM-Newton} and {\it INTEGRAL}, two ESA Science Missions with instruments and contributions directly funded by ESA member states and the USA (NASA). Also based on optical spectra collected at the Observatoire de Haute Provence (France).}}
\author{E.\ Mossoux\inst{1} \and J.~M.\ Pittard\inst{2} \and G.\ Rauw\inst{1} \and Y.\ Naz\'e\inst{1}\fnmsep\thanks{Senior Research Associate FRS-FNRS (Belgium).}}
\institute{Space sciences, Technologies and Astrophysics Research (STAR) Institute, Universit\'e de Li\`ege, All\'ee du 6 Ao\^ut, 19c, B\^at B5c, 4000 Li\`ege, Belgium
\and School of Physics and Astronomy, University of Leeds, Woodhouse Lane, Leeds LS2 9JT, UK} 
\date{}
\abstract{}
{\stars{} is a massive O-type binary displaying strong non-thermal radio emission. 
Owing to the compactness of this binary, emission of non-thermal X-ray photons via inverse Compton scattering is expected.} 
{We first revised the orbital solution for \stars{} using new optical spectra. 
We then reduced and analysed X-ray spectra obtained with {\it XMM-Newton}, {\it Swift}, {\it INTEGRAL,} and {\it NuSTAR}.}
{The analysis of the {\it XMM-Newton} and {\it Swift} data allows us to better characterise the X-ray emission from the stellar winds and colliding winds region at energies below 10\,keV. 
We confirm the variation of the broad-band light curve of \stars{} along the orbit with, for the first time, the observation of the maximum emission around phase 0.8. 
The minimum ratio of the X-ray to bolometric flux of \stars{} remains well above the level expected for single O-type stars, indicating that the colliding wind region is not disrupted during the periastron passage.
The analysis of the full set of publicly available {\it INTEGRAL} observations allows us to refine the upper limit on the non-thermal X-ray flux of the Cyg OB2 region between 20 and 200\,keV.
Two {\it NuSTAR} observations (phases 0.028 and 0.085) allow us to study the \stars{} spectrum up to 30\,keV.
These data do not provide evidence of the presence of non-thermal X-rays, but bring more stringent constraints on the flux of a putative non-thermal component.
Finally, we computed, thanks to a new dedicated model, the anisotropic inverse Compton emission generated in the wind shock region.
The theoretical non-thermal emission appears to be compatible with observational limits and the kinetic luminosity computed from these models is in good agreement with the unabsorbed flux observed below 10\,keV.
}
{}

\keywords{stars: early-type -- stars: massive -- binaries: spectroscopic -- stars: individual: Cyg~OB2~\#8a}
\authorrunning{E.\ Mossoux et al.}
\titlerunning{Cyg~OB2~\#8a}
\maketitle

\section{Introduction}
\label{intro}
The radiation fields of massive stars drive powerful stellar winds associated with huge mass-loss rates (typically $10^{-7}$ to $10^{-5}$\,M$_{\odot}$\,yr$^{-1}$) and highly supersonic velocities (typically $2000$\,km\,s$^{-1}$). 
In binary systems consisting of two massive stars, the collision of their stellar winds produces an interaction zone contained between two oppositely faced hydrodynamical shocks separated by a contact discontinuity \citep[e.g.][]{SBP}. 
The presence of this interaction zone leads to a number of observational signatures that can range from radio waves into the high-energy domain \citep[see e.g.][]{Rauw2013}. 
Thermal X-ray emission is created by the plasma from the interaction zone which may be heated up to $10^7$\,K through the kinetic energy of the winds \citep{SBP}. In some systems, this results in a prominent X-ray emission in the 0.5 -- 10\,keV band that varies with orbital phase, as a result of the changing optical depth along the line of sight and/or of the changing orbital separation in eccentric systems \citep[for a review, see][ and references therein]{RauwNaze}.

Aside from the heating of post-shock plasma, hydrodynamic shocks in colliding wind binaries (CWBs) can also produce a population of relativistic particles via diffusive shock acceleration through the first order Fermi mechanism. 
A subset of the CWBs indeed display synchrotron radio emission \citep[][and references therein]{Benaglia}, which is produced via the interaction of relativistic electrons with a magnetic field \citep{EU93,Pittard2006,PittDough}. 
This non-thermal radio emission is often variable as a result of changing line-of-sight optical depth and, in eccentric systems, also of changing shock strength \citep[][]{blomme10,Blomme2013}.

The presence of a population of relativistic electrons, along with the enormous supply of stellar photospheric ultraviolet (UV) photons, especially in short-period massive binaries, should result in a strong inverse Compton (IC) scattering emission in hard X-rays and soft $\gamma$-rays \citep{Pollock,CW,PittDough,Reimer,delPalacio}. 
X-ray observations of O-star binaries that display non-thermal radio emission however revealed no clear evidence of non-thermal X-ray emission in the 0.5 -- 10\,keV energy domain \citep{9Sgr,DeB06}. 
At these energies, the putative non-thermal X-ray emission is overwhelmed by the strong thermal emission from the wind interaction which involves plasma at temperatures up to $kT = 2$\,keV. 
However, at energies above 10\,keV, where the thermal emission rapidly declines, the conditions for detections of a non-thermal emission are more favourable.
Indeed, a detection of the very massive CWB $\eta$~Carinae above 20\,keV has been reported with {\it INTEGRAL} and {\it Suzaku}, and most recently with {\it NuSTAR} \citep{Leyder08,Leyder10,Sekiguchi,Hamaguchi}\footnote{{\it Suzaku} observations of the CWB WR~140 possibly also revealed such a hard X-ray component, although these measurements are likely contaminated by emission from the Seyfert~2 galaxy IGR~J20216+4359 \citep{Sugawara}.}.
The CWB was also detected up to giga-electronvolt energies with the {\it AGILE} and {\it Fermi} telescopes \citep{tavani09,abdo10}.
However, detection of the hard X-ray emission between 10 keV and 20\,keV with {\it NuSTAR} is somewhat uncertain \citep{Hamaguchi,panagiotou18}.

In the present work, we consider the CWB \stars{} ($\equiv$ BD+40$^{\circ}$~4227A, Schulte~8A), an eccentric ($e \sim 0.2$) O6\,I + O5.5\,III binary with an orbital period of 21.9\,days \citep{debecker04}.
It was one of the first massive stars found to be an X-ray emitter \citep{Harnden}. 
Since that pioneering detection, the system has been shown to display phase-locked variations in its X-ray emission \citep[][and references therein]{cazorla14}.
\stars{} is also known for its strong non-thermal radio emission \citep{Bieging,blomme10}. 
Owing to the compactness of its orbit, we expect IC scattering to be particularly efficient in this system.
\citet{INTEGRAL} analysed the {\it INTEGRAL}/IBIS data of the Cygnus region obtained during the two first Announcements of Opportunity, but failed to detect any emission directly associated with Cyg\,OB2.
These authors inferred 3$\sigma$ upper limits on the count rates at the position of the unidentified {\it EGRET} source 3EG\,2033+4118; these count rates, assuming a photon index $\Gamma = 1.5$, convert into upper limits on the flux of $6.1\,10^{-12}$, $4.2\,10^{-12}$, and $4.0\,10^{-11}\,\mathrm{erg\,s^{-1}\,cm^{-2}}$ in the 20 -- 60, 60 -- 100, and 100 -- 1000\,keV energy bands, respectively. 
In view of the spatial resolution of IBIS, these upper limits apply to the combined emission from all putative emitters within a radius of 6\arcmin{} around the position of 3EG\,2033+4118, which includes \stars{} as well as Cyg\,OB2~\#9, another non-thermal radio emitting O-star binary \citep[][and references therein]{HK, Blomme2013}.
In this work, we analyse the overall set of available {\it INTEGRAL} observations performed up to now and refine the upper limit on the hard X-ray emission from \stars{}.
The {\it NuSTAR} satellite is able to observe in hard X-rays with a much better angular resolution than {\it INTEGRAL}.
We also analyse the first {\it NuSTAR} observations of \stars{}.
We first present the observations (Sect.~\ref{obs}) and revise the orbital solution of the system (Sect.~\ref{orb_sol}).
We then present the X-ray spectral analysis (Sect.~\ref{spectre}) and discuss the results (Sect.~\ref{discuss}).
The summary and conclusions of this study are provided in Sect.~\ref{summary}.

\section{Observations}
\label{obs}
  
\subsection{\it NuSTAR}
\begin{figure}[t]
\centering
\includegraphics[trim= 15cm 3cm 15cm 0cm,clip, width=10cm]{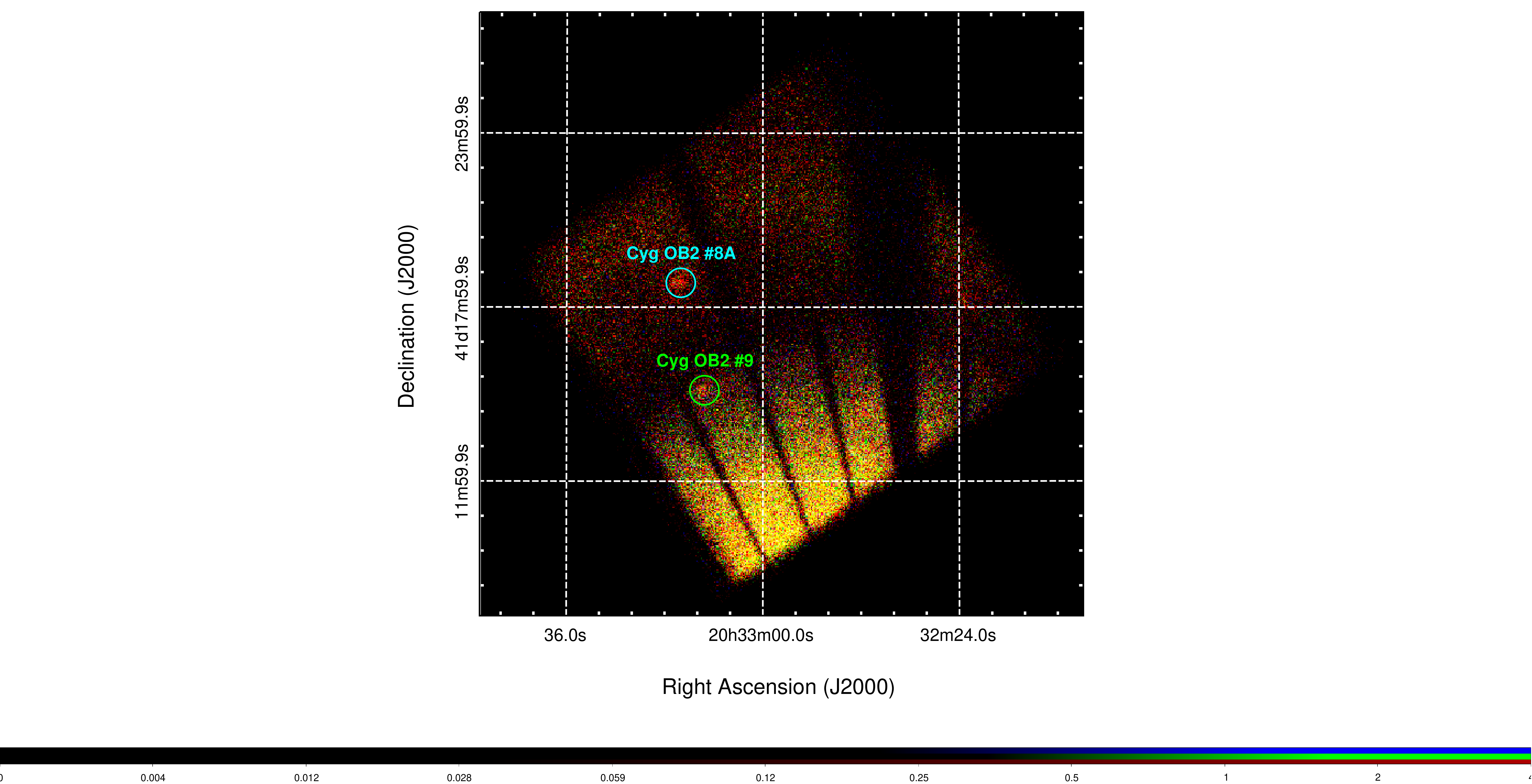}
\caption{{\it NuSTAR} RGB image (red=1.6--5.6$\,$keV, green=5.6--13.6$\,$keV, blue=13.6--21.6$\,$keV) of the Cyg~OB2 region taken on 2018 August 26 (FPMA data only).}
\label{figNustar}
\end{figure}

\stars{} was observed with the {\it NuSTAR} satellite on 25 and 26 August 2018. 
{\it NuSTAR} \citep{NuSTAR} features two focussing telescopes that operate in the energy domain from 3 keV to 79\,keV. 
The focussing optics provide a superior sensitivity in the hard X-ray domain compared to previous missions that relied on coded masks.
We analysed these observations with the {\it NuSTAR} Data Analysis Software package (\texttt{nustardas} v1.8.0) built in HEASOFT software (v$6.24$).
Stray light contamination from the off-axis source Cygnus~X-3 is high in the field of view of both observations (see Fig~\ref{figNustar}).
Moreover, part of the observations were taken close to the South Atlantic Anomaly (SAA).
Great caution must thus be taken to correctly take the high X-ray background into account.
The images reveal two sources: Cyg\,OB2~\#8a and Cyg\,OB2~\#9.
We focus on the former target because the latter source is weaker and is more heavily affected by the stray light from Cyg~X-3.

The events from \stars{} were extracted over a $30''$ radius circular area centred on the optical position of \stars{} \citep{gaia18}.
As recommended by the {\it NuSTAR} team\,\footnote{\href{https://heasarc.gsfc.nasa.gov/docs/nustar/analysis/nustardas_swguide_v1.7.pdf}{https://heasarc.gsfc.nasa.gov/docs/nustar/analysis/nustardas\_swguide\\ \_v1.7.pdf}}, the background region is defined as an annulus of 50 and $80''$ inner and outer radius centred on \stars{}.
Based on the configuration of the observations, we reduced the level-1 data using \texttt{nupipeline} (v$0.4.6$) with the keywords \texttt{tentacle=yes} and \texttt{saamode=strict}.
We then extracted the events and spectra from focal plane modules (FPMs) A and B for the source and background regions.
The corresponding response files (ARF and RMF) were created with \texttt{numkarf} and \texttt{numkrmf}.
The source spectra were grouped with a minimum signal-to-noise ratio of 4 in the \texttt{ISIS} software \citep{isis}.

  \subsection{\it XMM-Newton}
Twelve {\it XMM-Newton} observations were used in this study (Table~\ref{table_obs}).
Seven of these observations were previously analysed by \citet{cazorla14}.

For all observations but 0677980601, the European Photon Imaging Cameras \citep[EPIC; ][]{turner01,strueder01} were operated in full-frame mode. 
For observation 0677980601, the metal oxide semi-conductor (MOS) cameras were operated in large window mode with the field of view centred on Cyg~OB2~\#9.
\stars{} was thus only observed by the pn camera for this observation.
Given the optical brightness of our target ($V = 8.98$), the medium optical filter was used to reject optical and UV photons.
The EPIC data were processed with the \texttt{emchain} and \texttt{epchain} tasks from the Science Analysis Software (SAS) package (version 17.0; Current Calibration files as of 2018 June 22) to extract the event lists for the MOS and pn cameras.
We then selected the good time intervals (GTI) defined when the total count rate in the $0.2-10\,$keV energy range over the full detector is lower than $0.009$ for pn and $0.004\,\mathrm{count\,s^{-1}\,arcmin^{-2}}$ for MOS.
We selected the X-ray events by keeping only the single and double events ($\mathrm{\texttt{PATTERN}\leq4}$) for the pn camera and the single, double, triple, and quadruple events ($\mathrm{\texttt{PATTERN}\leq12}$) for the MOS cameras.
Finally, we rejected the dead columns and bad pixels using the bit masks \texttt{FLAG==0} for pn and \texttt{\#XMMEA\_SM} for MOS.
The source extraction region was the same as for the {\it NuSTAR} analysis.
This region allows us to extract 90 and 85\% of the flux at $1.5\,$keV on-axis for the pn and MOS cameras, respectively.
The background events were extracted from an annulus of 50 and $100''$ inner and outer radius centred on \stars{}.
The X-ray point sources detected in the background region using the SAS task \texttt{edetect\_chain} were filtered out.
We extracted the spectra for the source and background regions and built the corresponding ARF and RMF.
The source spectra were grouped with a minimum signal-to-noise ratio of 7 using the task \texttt{specgroup}. 

\subsection{\it Swift}
\stars{} was observed 73 times with {\it Swift} \citep{Gehrels} in photon counting mode from 2006 December 15 to 2018 August 27 (Table~\ref{table_obs}).
Six of these observations were previously analysed by \citet{cazorla14}.

We used the HEASOFT task \texttt{XRTPIPELINE} (v$0.13.4$) and the calibration files released on 2018 July 10 to reject the hot and bad pixels and select the grades between 0 and 12.
We used \texttt{XSelect} (v$2.4$e) to extract events from \stars{}. 
The definition of the source region is the same as for the {\it XMM-Newton} and {\it NuSTAR} data.
For the spectral analysis, the ARFs corresponding to the source position on the detector were computed with \texttt{xrtmkarf} (v$0.6.3$).
The spectra were grouped with a minimum of 20 counts per bin using \texttt{grppha}.
The spectra having less than three bins were rejected.

\subsection{\it INTEGRAL}
\label{integral}
{The \it INTEGRAL} instrument observed the Cygnus region thousands of times with the IBIS imager \citep{ubertini03}.
We selected all public observations performed with ISGRI up to now (i.e. from rev.\ 11 to 1977), where the Cygnus region is located in the fully coded field of view (i.e. with an off-axis angle lower than 480$\arcmin$), leading to 1736 observations corresponding to 3.1\,Ms of exposure time\footnote{These data were affected by ghost features from the very bright source Cyg~X-1 and to a lesser extent by other bright sources in the IBIS field of view.}.
We analysed these observations using the off-line scientific analysis (OSA) software (v11.0) provided by the Integral Science Data Center \citep{courvoisier03}.
We built a mosaic image of the overall {\it INTEGRAL} observations in the three following energy bands: 20--60, 60--100, and 100--200\,keV.
Following \citet{INTEGRAL}, we then extracted the spectra by considering a Gaussian with a half width at half maximum of 6$\arcmin$ at the position of the {\it EGRET} source 3EG~J2033+4118 corresponding to the Cygnus~OB2 region for each of the mosaic image with the \texttt{mosaic\_spec} task from the OSA software.
The response files and background spectra were created using the \texttt{spe\_pick} command to average the files proportionally to the exposures from the different observing periods.

\subsection{Optical spectroscopy}
Optical spectroscopy of \stars{} was obtained with the Aur\'elie spectrograph \citep{Gillet} at the 1.52\,m telescope of the Observatoire de Haute Provence (OHP; France). 
The data were taken during two observing campaigns of six nights each in June 2016 and September 2017. 
Aur\'elie was equipped with a $2048 \times 1024$\,CCD with a pixel size of 13.5\,$\mu$m squared. 
We used a 600 \,l/mm grating providing a reciprocal dispersion of 16\,\AA\,mm$^{-1}$. 
The resolving power, measured on the Thorium-Argon calibration exposures, was 7000 over the wavelength range from 4440 to 4890\,\AA.
Typical integration times were 1.5\,hour. 
The data were reduced using the {\sc midas} software (version 17FEBpl\,1.2).
In this way, we obtained eight new spectra (see Table\,\ref{journalvis}).

\section{Revised orbital solution}
\label{orb_sol}
We measured the radial velocities (RVs) of the He\,{\sc i} $\lambda$\,4471 absorption line on our new optical spectra. 
Because of the severe blending of the primary and secondary components at most orbital phases, the {\sc midas} two-Gaussian fit routine {\tt deblend/line} only provided a reliable fit for two observations (HJD~2\,458\,002.538 and HJD~2\,458\,004.516). 
We therefore adopted the same approach as \citet{debecker04} to measure the RVs on the other spectra: we fitted each spectrum with a combination of two Gaussians in which the widths and relative intensities were fixed to the values obtained via {\tt deblend/line} for the observation taken on HJD~2\,458\,002.538. 
In this procedure, only the positions of the lines and the overall line intensity were varied.
The resulting RVs are listed in Table\,\ref{journalvis}.
The typical errors on these new RVs are around 10\,km\,s$^{-1}$ for the primary star, but can exceed 20\,km\,s$^{-1}$ for the secondary at those phases where the lines are most heavily blended.

\begin{table}
  \caption{Journal of the new optical spectroscopy observations\label{journalvis}}
  \begin{center}
  \begin{tabular}{c c r r c}
    \hline
 
    HJD-2\,450\,000 & $\phi$ & \multicolumn{2}{c}{RV (km\,s$^{-1}$)} & Weight \\
    &        & \multicolumn{1}{c}{Primary} & \multicolumn{1}{c}{Secondary} & \\
    \hline
7547.497 & 0.085 &  $-40.8$ &   $66.5$ & 0.1 \\  
7548.470 & 0.130 &  $-25.4$ &   $33.6$ & 0.1 \\
7549.498 & 0.177 &  $-24.9$ &   $56.2$ & 0.1 \\
7551.569 & 0.271 &   $21.1$ & $-115.7$ & 0.1 \\
8002.538 & 0.857 &  $-81.3$ &   $87.6$ & 1.0 \\
8004.516 & 0.948 &  $-93.2$ &  $104.5$ & 1.0 \\
8005.555 & 0.995 &  $-98.9$ &   $69.4$ & 0.7 \\
8007.494 & 0.084 &  $-43.7$ &   $22.0$ & 0.1 \\
\hline
  \end{tabular}
  \end{center}
  \tablefoot{The last column lists the weight that we assigned to the RV measurements in the orbital solution.
  For consistency, we adopted the same definition of the weights as \citet{debecker04}: a weight of 1.0 for RVs from spectra taken at phases when the lines are well resolved down to 0.1 at phases when the lines are heavily blended.}
\end{table}

\begin{figure}[h]
\begin{center}
  \resizebox{8cm}{!}{\includegraphics{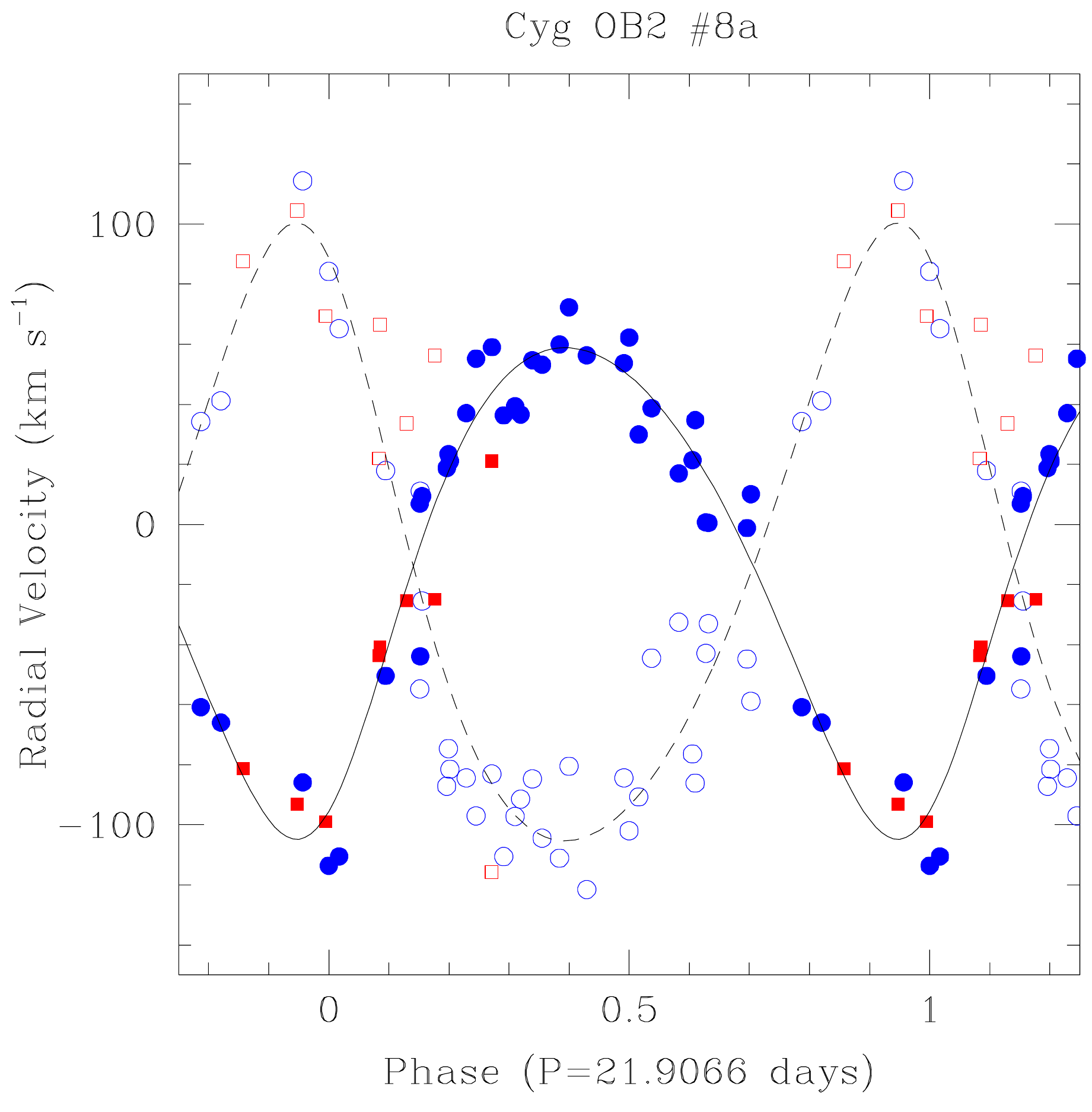}}
  \resizebox{8cm}{!}{\includegraphics{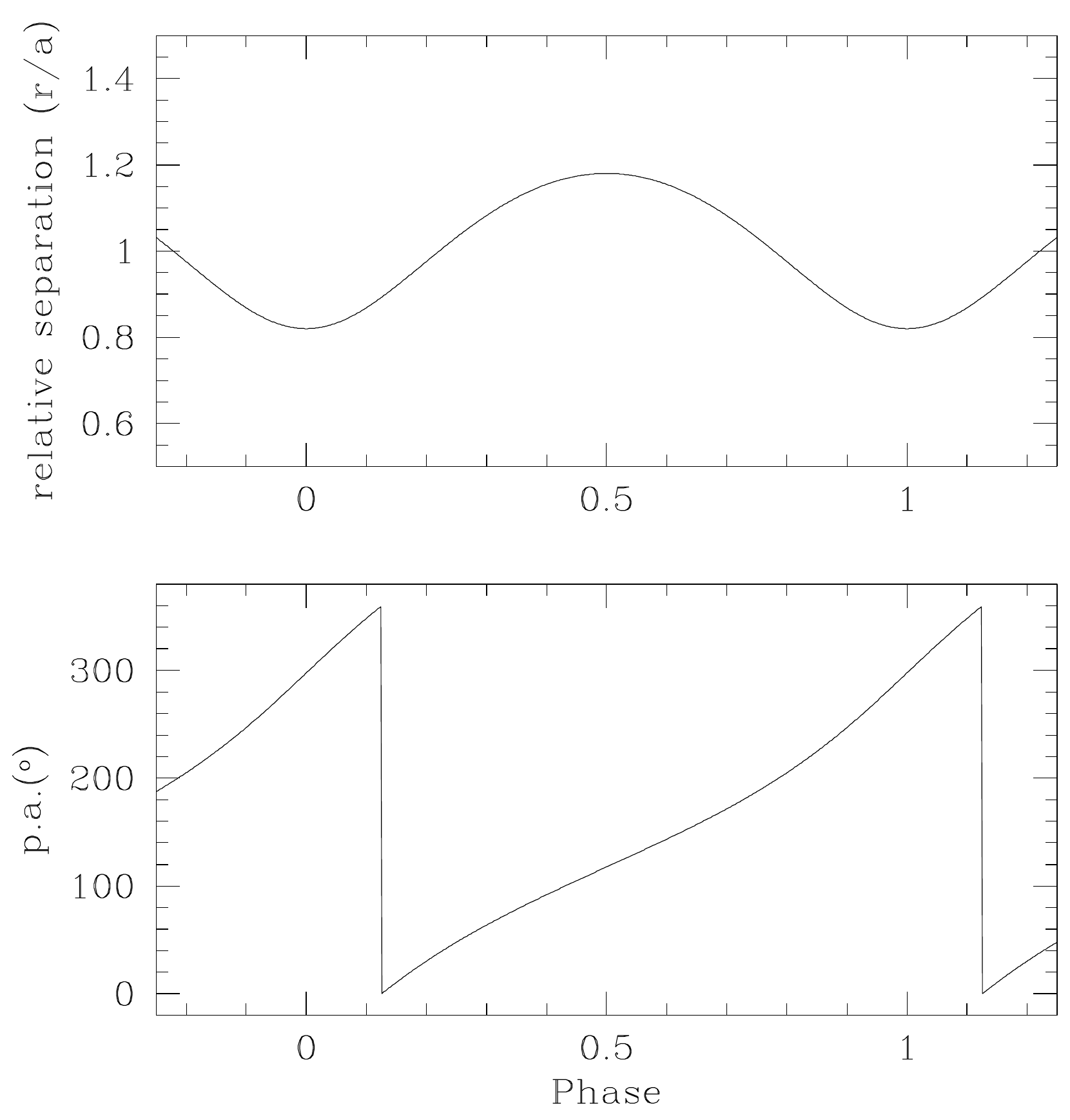}}
\end{center}  
\caption{Revised orbital solution of \stars{}. 
{\it Top panel:} The primary and secondary RVs are indicated by the filled and open symbols, respectively. 
The blue dots correspond to the RVs from \citet{debecker04}, whereas the red squares indicate the new RVs. 
The solid and dashed lines illustrate the best-fit RV curve (see Table\,\ref{solorb}). 
{\it Bottom panels:} Relative orbital separation and position angle as a function of orbital phase. A value of $0^{\circ}$ for the position angle corresponds to the conjunction with the primary star in front.}
\label{RVcurve}
\end{figure}

We then combined the new RV points with those of \citet{debecker04} to compute a revised orbital solution with the Li\`ege Orbital Solution Package (LOSP) code \citep{LOSP}. 
The results are given in Table\,\ref{solorb} and illustrated in Fig.\,\ref{RVcurve}. Compared to the solution of \citet{debecker04}, we find a lower eccentricity (0.18 vs.\ 0.24) and a slightly larger mass ratio (1.26 versus\ 1.16).  

\begin{table}
  \caption{Revised orbital solution of \stars{}. \label{solorb}}
  \begin{center}
\begin{tabular}{l c c}
  \hline
  & Primary & Secondary \\
  \hline
$P_{\rm orb}$ (days)     & \multicolumn{2}{c}{$21.9066 \pm 0.0013$} \\
$e$                    & \multicolumn{2}{c}{$0.18 \pm 0.03$} \\
  $T_0$ (HJD$-$2\,450\,000) & \multicolumn{2}{c}{$8005.66 \pm 0.62$} \\
  $\omega$ ($^{\circ}$)  & $207.6 \pm 11.9$ & \\
$\gamma$ (km\,s$^{-1}$) & $-10.0 \pm 2.8$ & $-18.8 \pm 3.2$ \\
  $K$ (km\,s$^{-1}$)      & $81.8 \pm 2.9$   & $102.8 \pm 3.6$ \\
  $q = m_1/m_2$         & \multicolumn{2}{c}{$1.26 \pm 0.06$} \\
$a\,\sin{i}$ (R$_{\odot}$) & $34.8 \pm 1.2$ & $43.8 \pm 1.5$ \\
$m\,\sin^3{i}$ (M$_{\odot}$) & $7.6 \pm 0.6$ & $6.0 \pm 0.5$ \\
rms (km\,s$^{-1}$)      & $13.8$ & $20.1$ \\
\hline
\end{tabular}
\end{center}
\tablefoot{$T_0$, $\omega$, $\gamma$, $K$, and $a\,\sin{i}$ stand, respectively, for the time of periastron passage, longitude of periastron of the primary measured from its ascending node, apparent systemic velocity, amplitude of the RV curve, and projected semi-major axis between the centre of the star and the centre of mass of the system.}
\end{table}

Comparing the minimum masses that we inferred from our revised orbital solution with the typical masses of stars of same spectral type quoted by \citet{Martins}, we estimated an orbital inclination of $(33.8 \pm 1.0)^{\circ}$ for the primary and $(32.9 \pm 1.0)^{\circ}$ for the secondary. 
We thus adopted $i = (33.4 \pm 0.7)^{\circ}$ as a reasonable estimate of the inclination. This then leads to a semi-major axis of the orbit of $a = 142.8\,$R$_{\odot} = 9.9\,10^{12}$\,cm.
\citet{bailer-jones18} computed a distance for \stars{} of $1.5\pm0.1$\,kpc.
With an optical brightness ratio (primary/secondary) close to 2 and the bolometric corrections corresponding to the spectral types derived by \citet{debecker04}, we obtained luminosities of $(7.2\pm0.1)\,10^5$\,L$_{\odot}$ for the primary star and $(4.3\pm0.1)\,10^5$\,L$_{\odot}$ for the secondary star. 
Using the spectral type -- effective temperature calibration of \citet{Martins}, these bolometric luminosities imply stellar radii of $(20.6\pm0.2)$\,R$_{\odot}$ for the primary and $(14.0\pm0.1)$\,R$_{\odot}$ for the secondary. 
Both stars therefore remain well inside their Roche lobes at all orbital phases.

\section{X-ray spectral analysis}
\label{spectre}
The goal of this study is to provide constraints on the properties of a putative hard non-thermal X-ray emission from \stars{}.
To do so, we analysed the {\it NuSTAR} and {\it INTEGRAL} spectra to test the presence of a power-law component at high energies. 
This requires a good characterisation of the X-ray spectrum at lower energies as    well because some of the thermal emission that dominates below 10\,keV could extend into the higher energy domain. 
All our spectral fits were done under the X-ray spectral fitting package (XSPEC; v$12.10.0e$).

The X-ray spectrum of \stars{} is heavily absorbed at energies below 0.8\,keV by the large column density of interstellar material and by the material of the stellar winds. 
For the interstellar medium, we followed \citet{cazorla14} in adopting a total neutral hydrogen column density of $N_{\rm H} = 0.91\,10^{22}$\,cm$^{-2}$ corresponding to the \citet{Bohlin} relation with $E(B - V) = 1.56$. 
This value is higher than the $N_{\rm H\,\textsc{i}}$ value of $(0.63 \pm 0.07)\,10^{22}$\,cm$^{-2}$ inferred by \citet{Herrero} from the interstellar Ly\,$\alpha$ line measured on {\it HST/STIS} spectra of \stars{}. 
However, the Ly\,$\alpha$ line only traces H\,{\sc i}, whereas the total neutral hydrogen column density must also account for H$_2$. 
Assuming that the average $N_{\rm H_2}/N_{\rm H\,\textsc{i}}$ ratio of \citet{Bohlin} holds for the sightline towards \stars{}, the results of \citet{Herrero} translate into $N_{\rm H} = (0.84 \pm 0.10)\,10^{22}$\,cm$^{-2}$ in reasonable agreement with the value we adopted.

\subsection{Improvement of the spectral model of Cazorla et al.}
\label{spectral_model}
\citet{cazorla14} fitted seven of the {\it XMM-Newton} spectra with an absorbed optically thin thermal plasma model consisting of three components produced by the astrophysical plasma emission code \citep[APEC,][]{smith01}.
The parameters of each APEC component are the temperature of the plasma ($kT$) and the normalisation of the emission ($n$).
The three APEC models are absorbed by both the interstellar medium reproduced by \texttt{wabs} and the stellar winds modelled by \texttt{phabs}.

The \texttt{wabs} model uses the abundances determined by \citet{anders89}.
However, these abundances were updated by \citet{wilms00} leading to lower metal abundances.
We thus improved the \citet{cazorla14} model with the \texttt{TBnew} model, the abundances of \citet{wilms00}, and the updated cross sections of \citet{verner96}.
These abundances were also used for \texttt{phabs} and the APEC emission.
We also took the effects of the dust scattering along the line of sight into account by adding the \texttt{dustscat} component \citep{predehl95} whose hydrogen column density was set to the hydrogen column density of the \texttt{TBnew} component divided by 1.5 \citep{nowak12}.

We fitted the spectra using the Markov chain Monte Carlo (MCMC) method.
We used the {\tt XSPEC\_emcee}\footnote{\href{https://github.com/jeremysanders/xspec\_emcee}{https://github.com/jeremysanders/xspec\_emcee}} programme designed by Jeremy Sanders, which allows the MCMC analysis of X-ray spectra in {\tt XSPEC} using {\tt emcee}\footnote{\href{http://dan.iel.fm/emcee/current/user/line/}{http://dan.iel.fm/emcee/current/user/line/}}, an extensible, pure-Python implementation of the \citet{goodman10} MCMC ensemble sampler.
This method uses a number of `walkers', which evolve independently from each other in the parameter space reducing the autocorrelation time.
We followed the same methodology as adopted by \citet{cazorla14}. We first fitted the {\it XMM-Newton} spectra for each observation and checked the constancy of the temperatures of the three APEC components with phase.
Since they agree within the errors, we then set the temperatures to their mean values (0.34, 0.83 and 2.13\,keV) and fitted the {\it XMM-Newton} spectra again to determine the normalisation of the APEC models.
The normalisation of the softest component (which is probably mainly due to X-ray emission from the individual stars) is almost constant with phase and has very large error bars.
We thus fixed it to its mean value: $2.11\, 10^{-2}\,\mathrm{cm^{-5}}$.
We finally fitted {\it XMM-Newton} spectra adopting as free parameters the \texttt{phabs} hydrogen column density and the normalisations $n_2$ and $n_3$.

\begin{figure}[t]
\centering
\includegraphics[trim= 0cm 0cm 0cm 1.5cm,clip, width=9cm]{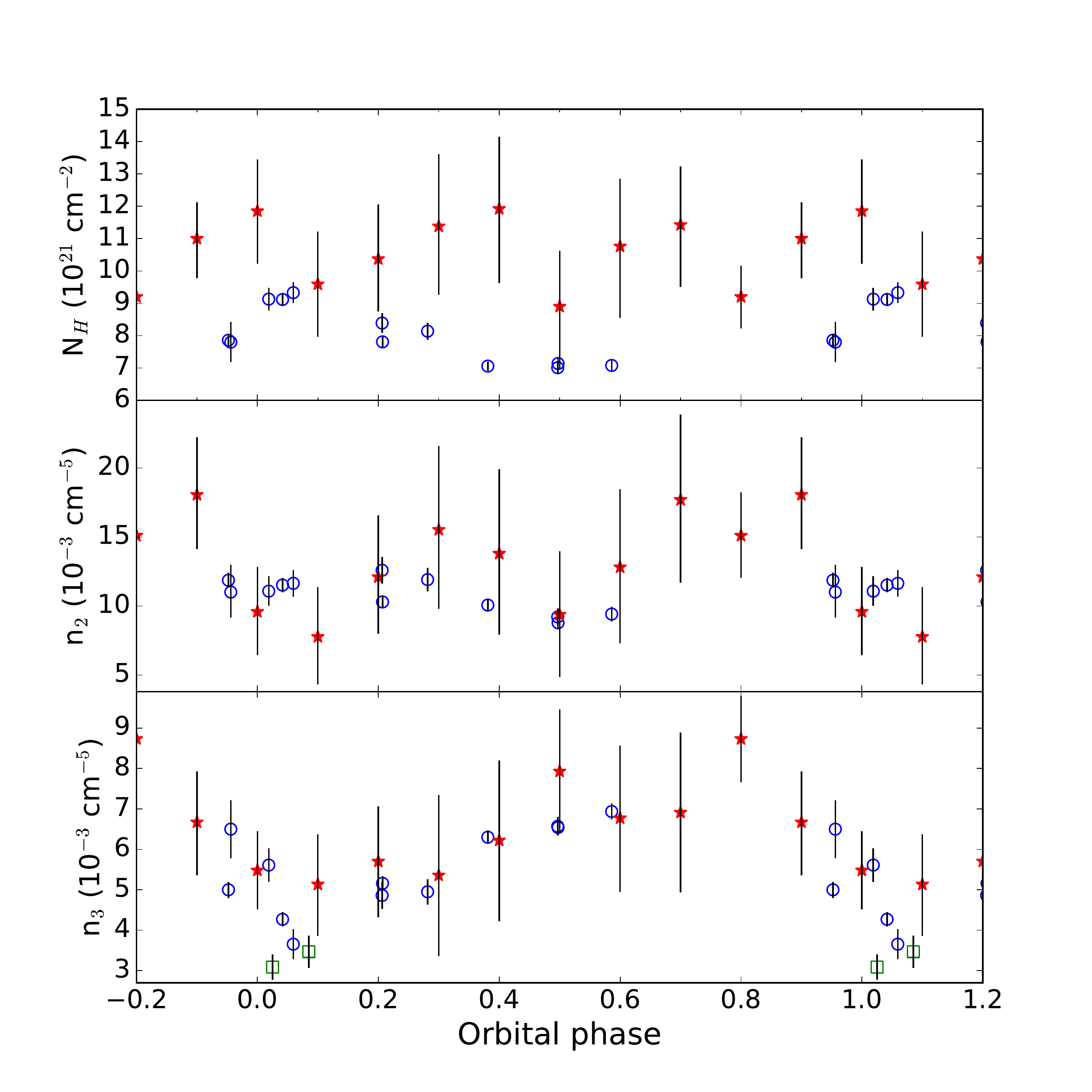}
\caption{Evolution with phase of the absorption and normalisation factors from the fit with three APEC components whose temperatures and $n_1$ component were fixed.
The red asterisks, blue circles, and green squares indicate the spectral parameters from the {\it Swift}, {\it XMM-Newton,} and {\it NuSTAR} fits, respectively.}
\label{figPar2}
\end{figure}

We also grouped the {\it Swift} spectra by phase interval; we used an interval of 0.1 beginning at $\phi=0.05$.
We simultaneously fitted the {\it Swift} spectra from each phase group with the same model as for {\it XMM-Newton}.
The variation in the spectral parameters with phase is shown in Fig.~\ref{figPar2}.
The best-fitting parameters of all the models we tested are summarised in Table~\ref{table_fit}. 

\subsection{Search for a non-thermal hard X-ray component}
\label{non-therm}
\begin{figure}[t]
\centering
\includegraphics[trim= 0cm 0cm 0cm 0cm,clip, width=8.8cm,angle=0]{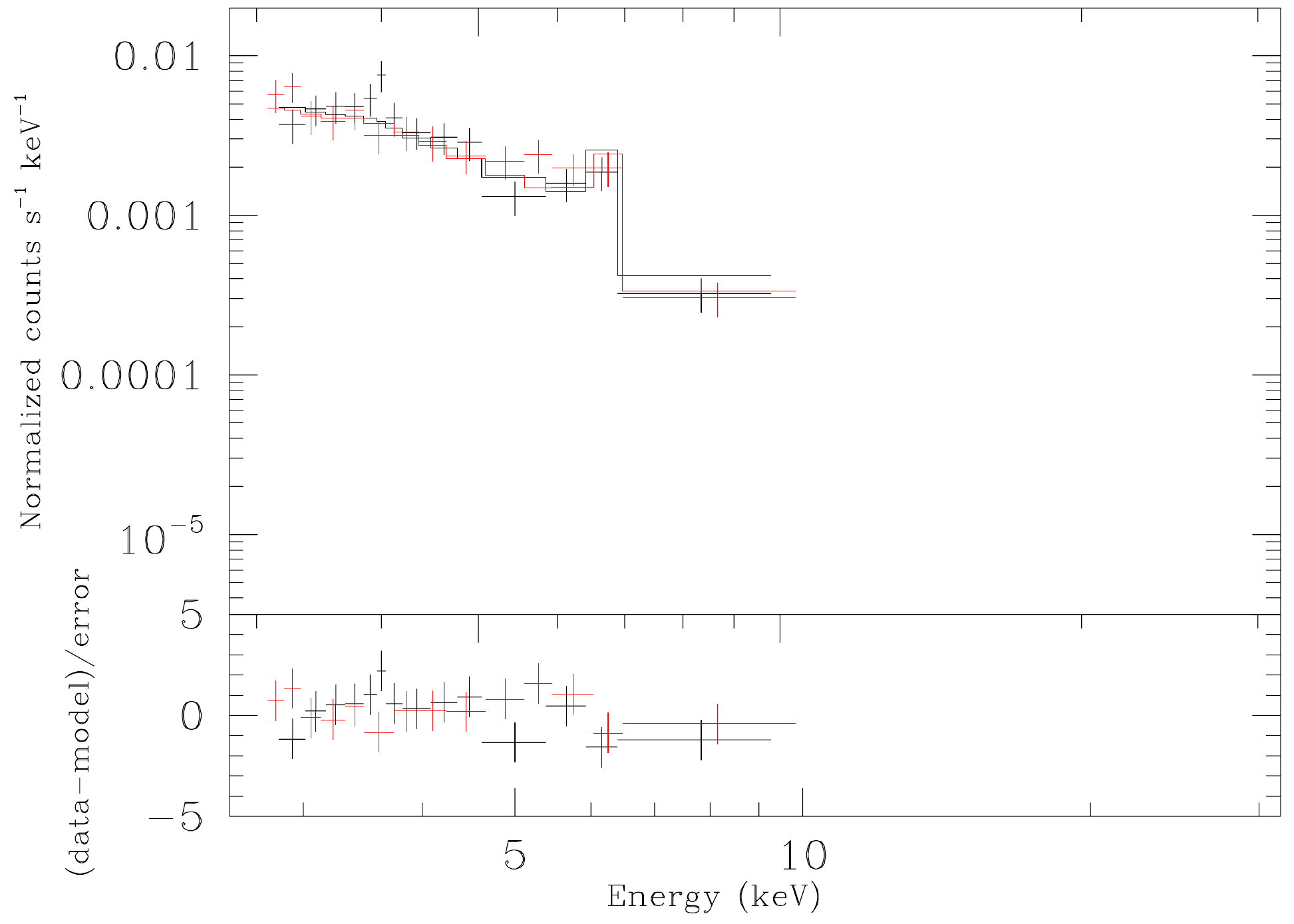}

\includegraphics[trim= 0cm 0cm 0cm 0cm,clip, width=8.8cm,angle=0]{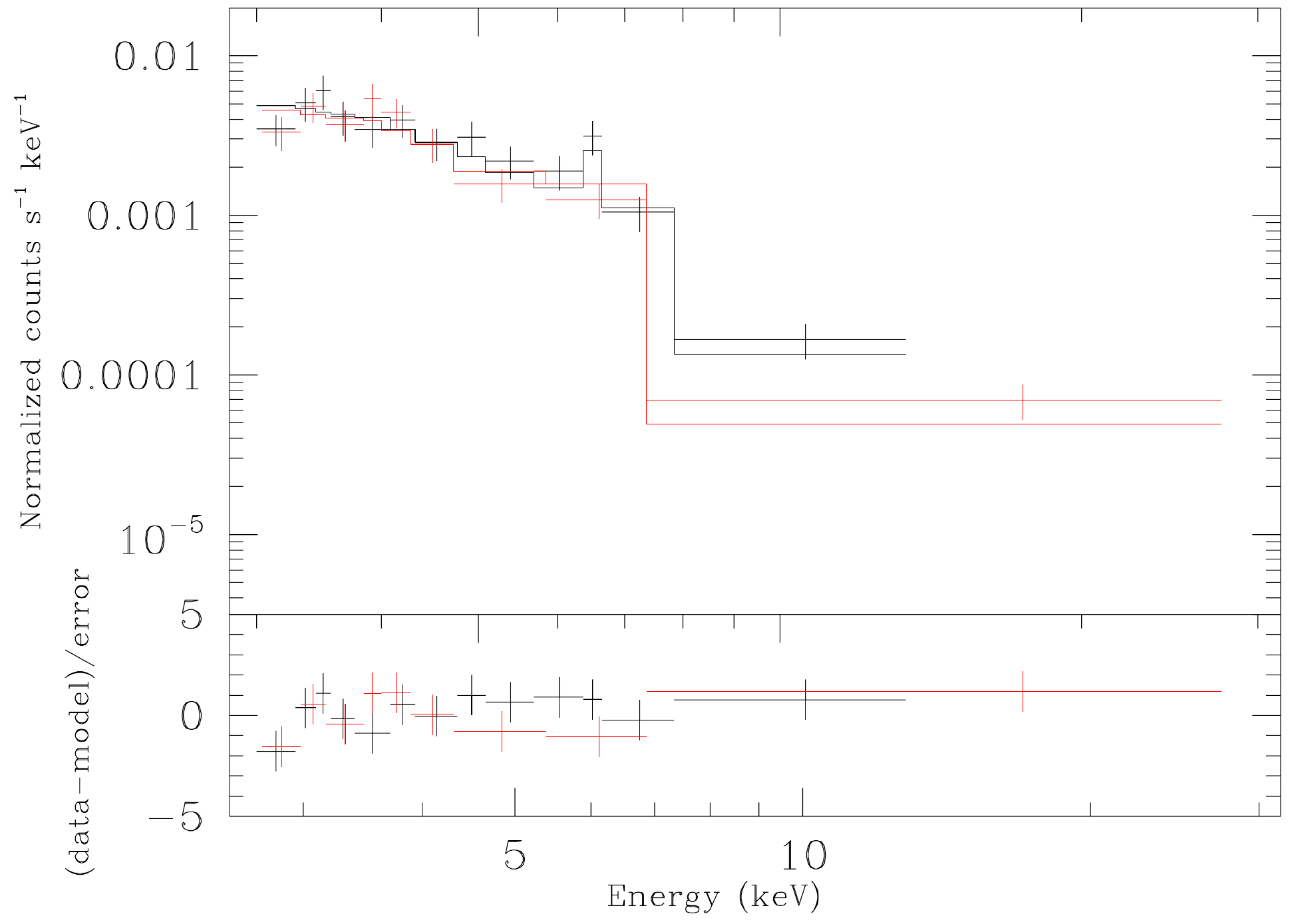}
\caption{Best-fitting dustscat$\times$TBnew$\times$phabs$\times$(apec+apec+apec) models of the $\phi=0.028$ (top panel) and $\phi=0.085$ (bottom panel) {\it NuSTAR} spectra.}
\label{figModel}
\end{figure}

The putative non-thermal photons emitted by the wind shock region mainly contribute to the flux observed at high energies, i.e. above 10\,keV.
To test the presence of such a component, we thus investigated the {\it NuSTAR} spectra.
The two {\it NuSTAR} observations were taken close to periastron passage (phase 0). 
We thus interpolated the values of the wind hydrogen column density and the $n_2$ and $n_3$ components to the phases of the {\it NuSTAR} observations and applied the so-constructed models to the {\it NuSTAR} spectra.
For the $\phi=0.028$ spectra, the interpolated model predicts a hard X-ray flux exceeding what is observed with {\it NuSTAR}.
We thus let the $n_3$ component free, leading to a best-fit value of $n_3=(3.3\pm0.3)\, 10^{-3}\,\mathrm{cm^{-5}}$ (Table~\ref{table_fit} and top panel of Fig.~\ref{figModel}).
This value is small compared to what is observed at phases 0.019 and 0.042 with {\it XMM-Newton}.
On the other hand, for the $\phi=0.085$ {\it NuSTAR} spectra, the observed flux above 10\,keV is underestimated by the interpolated model.
We thus added a power-law component to the model and fitted {\it NuSTAR} spectra letting only the power-law parameters free.
This leads to a negative photon index and a flux consistent with zero.
The F-statistic shows that the addition of the power-law component does not significantly improve the fit ($p=0.25$).
We thus let the $n_3$ component of the thermal model free leading to a best fit of $n_3=(3.7\pm0.4)\, 10^{-3}\,\mathrm{cm^{-5}}$, i.e. close to the value observed at phase 0.056 with {\it XMM-Newton} (bottom panel of Fig.~\ref{figModel}).

\subsection{Upper limits on the flux of a non-thermal component}
\label{upperlimits}
As becomes obvious from the various trials above, there is currently no unambiguous detection of a hard, non-thermal X-ray emission in \stars{}. 
We can nevertheless use the {\it INTEGRAL} and {\it NuSTAR} spectra to establish upper limits on the flux of such a component.

Using the spectra extracted from the mosaic images of the 3.1\,Ms of {\it INTEGRAL} observations, we constrained the 3$\sigma$ upper limit on the count rate of the 3EG\,2033+4118 source in the three energy bands to $3.5\,10^{-2}\,\mathrm{count\,s^{-1}}$ in 20--60\,keV, $6.3\,10^{-3}\,\mathrm{count\,s^{-1}}$ in 60--100\,keV, and $4.4\,10^{-3}\,\mathrm{count\,s^{-1}}$ in 100--200\,keV.
We then determined the flux in the 20--30\,keV energy band reproducing the total count rate of $4.5\,10^{-2}\,\mathrm{count\,s^{-1}}$ over the 20--200\,keV energy band.
Because the hard X-rays in this band should mostly stem from IC scattering, the total count rate was converted to flux by modelling the emission with a power-law characterised by different values of spectral index $\Gamma$.
The resulting 3$\sigma$ upper limits on the non-thermal flux of 3EG\,2033+4118 in the 20--30\,keV energy band are reported in Table~\ref{table_integral} for each value of $\Gamma$.
As mentioned in Sect.~\ref{intro}, these fluxes contain the emission from all putative emitters within a radius of 6$\arcmin{}$ around the position of 3EG\,2033+4118.
They are thus very large upper limits for the non-thermal emission from \stars{}.

We then deduced directly the 3$\sigma$ upper limit on the non-thermal flux from \stars{} observed during the two {\it NuSTAR} observations in the same energy band (20--30\,keV).
To do so, we applied an absorbed 3T plasma plus power-law model on the four {\it NuSTAR} spectra.
We constructed the thermal part of the model with the best-fitting parameters determined in Sect.~\ref{non-therm} for each spectrum (see Table~\ref{table_fit}).
The non-thermal part of the model is constructed with a power-law whose spectral index is fixed to the different values of $\Gamma$ reported in Table~\ref{table_integral}.
For each value of $\Gamma$, we determined a 3$\sigma$ upper limit on the flux of the power-law in the 20--30\,keV energy band.
These upper limits (also reported in Table~\ref{table_integral}) are more stringent than those determined with {\it INTEGRAL} observations.
In particular, for the value $\Gamma=1.5$ prevailing for the IC emission, the upper limit on the non-thermal emission is twice better when using {\it NuSTAR} observations.

\begin{table}
  \caption{Flux upper limits of the non-thermal emission of \stars{} as a function of $\Gamma$. \label{table_integral}}
  \begin{center}
\begin{tabular}{c c c}
  \hline
  $\Gamma$ & $F_\mathrm{INTEGRAL}$ & $F_\mathrm{NuSTAR}$\\
  & ($10^{-14}\,\mathrm{erg\,s^{-1}\,cm^{-2}}$) & ($10^{-14}\,\mathrm{erg\,s^{-1}\,cm^{-2}}$) \\
  \hline
1.0& 7.8& 7.1 \\
1.1& 8.4& 6.5\\
1.2& 9.0& 6.0\\
1.3& 9.5& 5.5\\
1.4& 10.1& 5.0\\
1.5& 10.6& 4.6\\
1.6& 11.1& 4.1\\
1.7& 11.6& 3.7\\
1.8& 12.0& 3.3\\
1.9& 12.4& 3.0\\
2.0& 12.7& 2.6\\
2.1& 13.0& 2.4\\
2.2& 13.3& 2.1\\
2.3& 13.5& 1.9\\
2.4& 13.7& 1.6\\
2.5& 13.8& 1.5\\
\hline
\end{tabular}
  \end{center}
  \tablefoot{$F_\mathrm{INTEGRAL}$ and $F_\mathrm{NuSTAR}$ stand for the 3$\sigma$ upper limits in the 20--30\,keV energy band on the flux of a non-thermal component as allowed by {\it INTEGRAL} observations of the Cygnus~OB2 region and {\it NuSTAR} observations of \stars{}}
\end{table}

\section{Discussion}
\label{discuss}

\subsection{Light curve in 2 -- 10\,keV }
\begin{figure}[t]
\centering
\includegraphics[trim= 0cm 0cm 0cm 0cm,clip, width=4cm,angle=90]{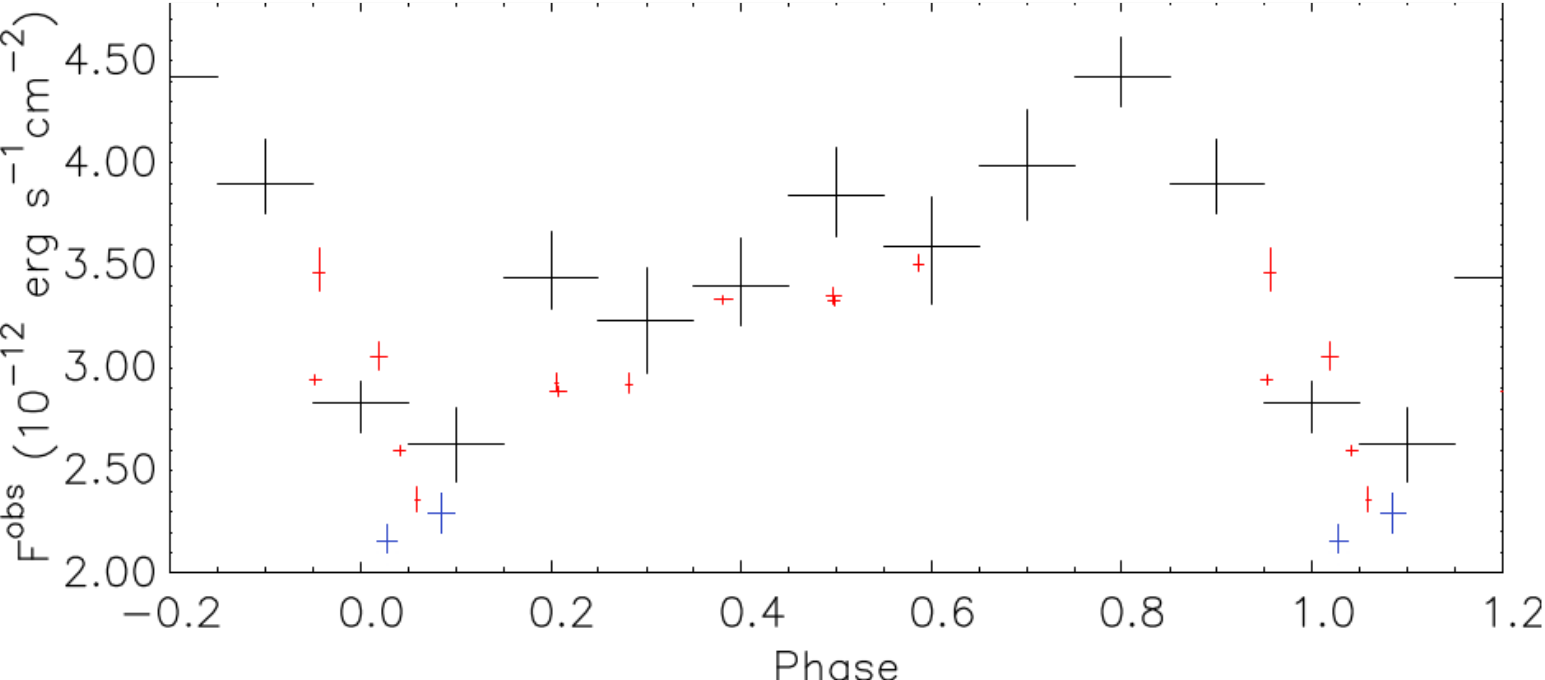}
\caption{\stars{} light curve in the $2-10\,$keV band. The black, red, and blue lines indicate the observed flux and error bars for the {\it Swift}, {\it XMM-Newton,} and {\it NuSTAR} spectra.}
\label{lcFlux}
\end{figure}

From the best-fitting models presented in Sect.~\ref{spectral_model}, we computed the \stars{} light curve showing the evolution of its observed (absorbed) flux over the entire orbit (Fig.~\ref{lcFlux}). 
This new X-ray light curve is more densely sampled than in the previous studies of \citet{DeB06}, \citet{blomme10}, \citet{Yoshida}, and \citet{cazorla14}. 
This allows us to get further insight into the properties of the wind interaction. 
Our light curve clearly reveals a maximum emission near $\phi \sim 0.8$, whilst the minimum occurs shortly after periastron \citep{DeB06,cazorla14}. 
Our spectral fits indicate that the emission measure of the hardest plasma component follows the same trend as the global flux.

Among the CWBs that have been monitored in detail in X-rays, the system that comes closest to \stars{} is probably HD~166\,734 \citep{HD166734}. 
This system consists of two O-type supergiants (O7.5\,I + O9\,I) on a highly eccentric orbit ($e = 0.62$) with a period of 34.5\,days \citep{Mahy}. 
The light curve of this system displays a minimum around the periastron which is attributed to a disruption of the shock \citep{HD166734}. 
Indeed, the residual X-ray emission of HD~166\,734 at this phase range is consistent with purely intrinsic emission from the two supergiants.
In \stars{}, the situation is different: even at minimum, the system remains overluminous in  X-rays by about a factor 20 ($\log(\mathrm{L_X/L_{bol}})=-5.4$), indicating that the shock does not get disrupted at periastron.

Finally, we note that the shape of the observed light curve in the 2--10\,keV domain is very close to the model cwb4 computed by \citet{Pittard2010}.
The authors modelled the wind shock region in 3D taking the Coriolis deflection and the radiative cooling into account.
They considered two O6V stars with mass-loss rates of $2\,10^{-7}$\,M$_{\odot}$\,yr$^{-1}$ and terminal velocities of 2500\,$\mathrm{km\,s^{-1}}$ orbiting each other with a period of 6.1~days and an eccentricity of 0.36.
This eccentricity is higher than that measured for \stars{} explaining the larger amplitude of the variations in the cwb4 light curve.
The decay of the light curve of cwb4 near the periastron is explained by a change in the cooling regime of the wind shock.
Indeed, close to periastron, the shock, which is adiabatic over most parts of the orbit, becomes radiative leading to a sudden cooling of the shocked gas.
A similar process may thus explain the shape of the light curve of Cyg OB2 \#8a.

\begin{figure}[t]
\begin{center}
\includegraphics[trim= 12.4cm 0cm 2.1cm 0cm,clip, width=7cm]{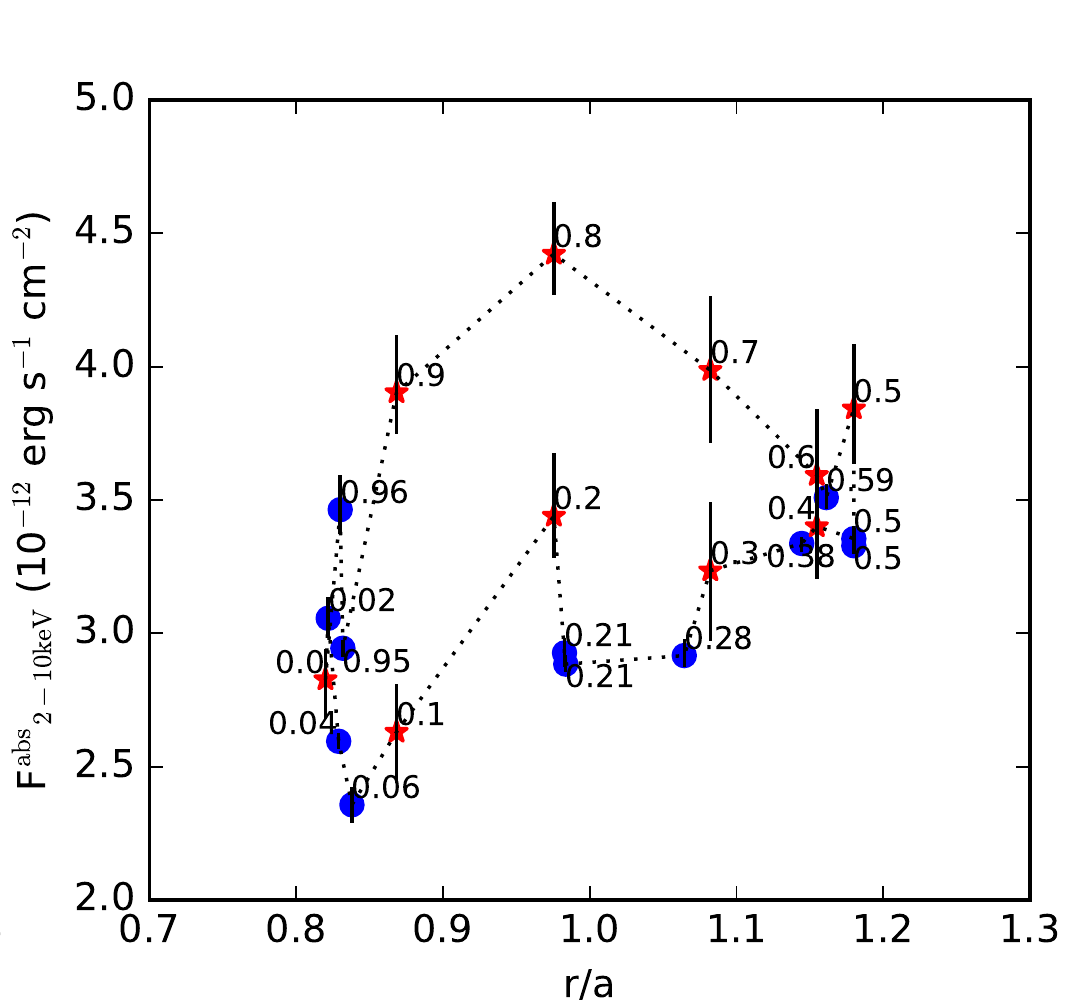}
\end{center}  
\caption{Observed X-ray flux of \stars{} in the 2 -- 10\,keV energy bands as a function of orbital separation computed with the revised orbital solution (see Table\,\ref{solorb}). 
Blue filled circles correspond to {\it XMM-Newton} spectra, whilst red asterisks stand for {\it Swift} observations binned in orbital phase (0.1 phase bins).}
\label{hysteresisX}
\end{figure}

Another result that can be noticed is the complexity of the dependence of the observed (absorbed) X-ray flux on orbital separation.
Indeed, the flux versus orbital separation plot describes a strong hysteresis \citep[see Fig.\,\ref{hysteresisX} and][]{cazorla14}, as theoretically predicted by \citet{PittPark}: at the same orbital separation, the emission of \stars{} is different depending on whether the primary is approaching or receding from the secondary.

\subsection{Evolution of the wind column density}
The variations in the X-ray light curves of CWBs can be due to either phase-dependent, line-of-sight wind column densities as the stars move around each other and/or to the variations in the orbital separation with phase. 
In view of the relatively low orbital inclination inferred in Sect.~\ref{orb_sol}, we expect the strongest influence to come from the changing orbital separation. 
Nevertheless, we observe some variations in the best-fit wind column density, in which the wind absorption is stronger at phases near periastron than at apastron (see top panel of Fig.~\ref{figPar2}). 
This situation most likely reflects the fact that the X-ray emitting part of the wind interaction zone is somewhat compact (i.e. dominated by the regions near the shock apex), is more deeply buried inside the winds at periastron, and is seen through the denser winds at that time.
Independent pieces of evidence that the X-ray emission zone is somewhat compact comes from a {\it Suzaku} observation analysed by \citet{Yoshida}. 
These authors found the flux above 3\,keV to vary on timescales of 20\,ks. 
In the 3T best-fit model of \citet{Yoshida}, this modulation reflects temperature changes (at the 20\% level) of the hottest plasma component. 
Interpreting these variations as hydrodynamical instabilities of the wind interaction zone, \citet{Yoshida} estimated a volume of about $2.4\,10^{36}$\,cm$^3$ for the hottest plasma, corresponding to a radius (for a spherical plasma volume) of $8.3\,10^{11}$\,cm = 11.9\,R$_{\odot }$ (i.e. 0.1 times the orbital separation at periastron). 
The actual shape of the wind contact surface is closer to a plane or a cone, and the hottest plasma is likely confined to a layer close to this surface. 
Therefore, the size of the X-ray emitting region is most probably larger than the stellar radii, but somewhat smaller than the orbital separation, consistent with theoretical expectations.

Whilst no clear trend can be detected in the absorptions fitted to the noisier {\it Swift} data, the results obtained for the {\it XMM-Newton} spectra indicate a much clearer trend (Fig.\,\ref{figPar2}). 
The {\it XMM-Newton} column density is largest at phases shortly after periastron when the primary is in front.
If the shock cone is turned towards the secondary, the denser primary wind would be in front of the X-ray emitting plasma near the shock apex at phases between 0.95 and 0.39 (see bottom panel of Fig.\,\ref{RVcurve}), which is consistent with the observations.

We adapted the model of \citet{wil90} to evaluate the line-of-sight column density as a function of orbital phase and attempt to constrain the orbital inclination. 
However, whereas the trend with phase appears well reproduced, the inclination $i = \left(20^{+6}_{-4}\right)$\,$^{\circ}$ fitted with this model is significantly lower than our estimate based on the typical masses of the stars, and may be explained by the limitations of this model (see appendix~\ref{williams_model}).  

\subsection{Predictions on the inverse Compton emission}
Before we can address the issue of the presence or absence of an IC hard X-ray emission, it is important to review the evidence for the presence of non-thermal electrons in the winds of \stars{}.
The radio emission of \stars{} is thought to arise from a synchrotron emission process because the emission level is well above that expected from free-free processes in the wind of the stars given their wind density.
Moreover, the analysis of the centimetre and millimetre emission shows phase-locked variations clearly revealing that a significant part of this emission arises from the wind interaction region \citep{blomme10,Blomme2017}. 
The thermal free-free emission of the hot high-density gas of the colliding wind region may also give rise to phase-locked variability in excess compared to that produced by the winds of the individual stars.
However, the predicted spectral indices in the centimetre wavelength domain for such a purely thermal emission are significantly positive \citep{Pittard2010} and would thus be difficult to reconcile with the observed spectral index which was found to be close to 0 \citep{blomme10}.
The radio emission of \stars{} is thus most probably emitted by the non-thermal radiation from relativistic electrons.

In a relatively close CWB such as \stars{}, a significant population of relativistic electrons is required to produce an observable synchrotron radio emission despite the large optical depths of the winds in the radio domain \citep{blomme10}.
In the innermost parts of the binary these electrons should suffer a substantial loss in energy via IC scattering of the intense stellar UV radiation field.
The corresponding emission should take the form of a hard X-ray emission with a power-law spectral energy distribution \citep{PittDough}.

The energy of IC scattered photons is given by
\begin{equation}
  h\,\nu_{\rm IC} \simeq \frac{4}{3}\,\gamma^2\,h\,\nu_*
,\end{equation}
where $\gamma$ stands for the Lorentz factor of the relativistic electrons. 
\citet{Blomme2017} showed that electron energies of about $450$\,MeV, corresponding to $\gamma \simeq 880$, are required to explain the 3\,mm emission via the synchrotron mechanism. 
Adopting the spectral type--effective temperature calibration of \citet{Martins}, we find that the spectra of the two stars in \stars{} peak near 16\,eV.
Therefore, adopting a conservative value for the Lorentz factor of the relativistic electrons of the order of 100, we expect the IC emission to extend out to energies of at least 100\,keV.  

To predict the flux of the IC emission from \stars{} we used the theoretical model of the wind-wind collision region and the resulting non-thermal emission created by \citet[][submitted]{pittard20}. 
The main characteristics of this model are summarised here.
The non-thermal particle distribution is calculated assuming that the shocks are coincident with the contact discontinuity. 
The immediate post-shock distribution of the primary particles (i.e. directly accelerated at the shocks) is computed at each point along the shocks using the semi-analytic model of \citet{blasi05}.
The diffusive shock acceleration leads to a spectral index of the non-thermal particle distribution that can be energy dependent.
The non-thermal particle distribution is then evolved downstream of the shocks by solving the kinetic equation. 
Energy losses due to IC, synchrotron and relativistic bremsstrahlung emission, and ionic and adiabatic cooling are calculated assuming a black-body distribution of incoming stellar photons and electron to proton number density ratio of 0.01.

\begin{figure}[t]
\begin{center}
\includegraphics[width=10cm]{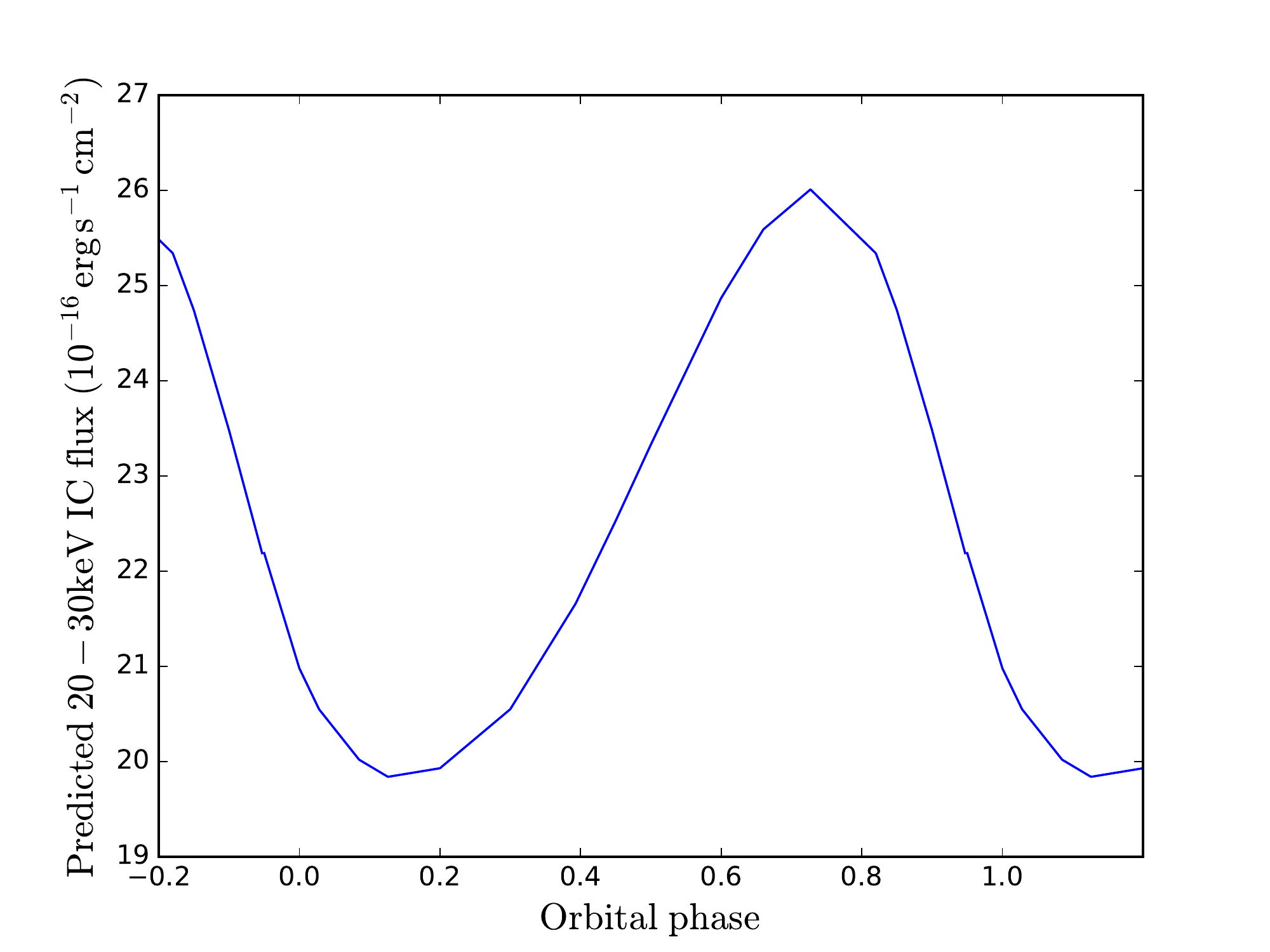}
\includegraphics[width=9.5cm]{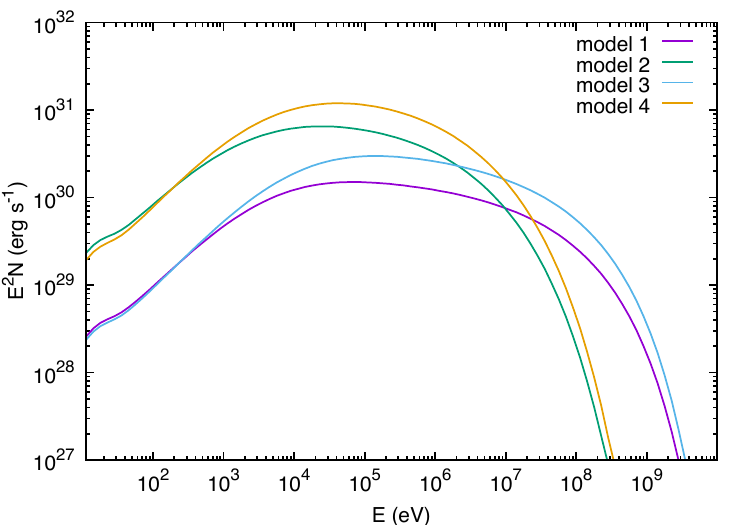}
\end{center}  
\caption{Predicted anistotropic IC emission from \stars{}.
{\it Top panel:} Model 1 along the orbital phase.
{\it Bottom panel:} Spectral energy distribution of model 1 (purple), 2 (green), 3 (blue), and 4 (orange) at orbital phase 0.028.}
\label{nt}
\end{figure}

We created several models differentiating by the values of the mass-loss rates $\dot{M}$ (varying the wind velocities has less impact on the value of the predicted IC emission) and pre-shock magnetic field strength $\zeta_{\rm B}$ which is very uncertain.
The last parameter is defined as the ratio between the pre-shock magnetic and the kinetic energy densities. 
Table~\ref{models} lists these values for the four models presented in Fig.~\ref{nt}.
\begin{table}
  \caption{Model parameters for four predictions of the anistotropic IC emission. \label{models}}
  \begin{center}
\begin{tabular}{c c c c}
  \hline
  Models & $\zeta_\mathrm{B}$ & $\dot{M}$ primary & $\dot{M}$ secondary\\
  & & (M$_{\odot}$\,yr$^{-1}$) & (M$_{\odot}$\,yr$^{-1}$)\\
  \hline
Model 1& $10^{-3}$& $10^{-6}$ & $10^{-7}$\\
Model 2& $10^{-6}$& $10^{-6}$ & $10^{-7}$ \\
Model 3& $10^{-3}$& $8\,10^{-6}$ & $2\,10^{-7}$ \\
Model 4& $10^{-6}$& $8\,10^{-6}$ & $2\,10^{-7}$ \\
\hline
\end{tabular}
  \end{center}
\end{table}

The top panel of Fig~\ref{nt} shows the predicted anisotropic IC emission along the orbital phase for model 1. 
The predicted IC emission is maximum prior to periastron and decreases by about 25\% near phase 0.2. 
The variation is similar for the three other models.
It is caused by the changing separation of the stars and the orientation of the stars and the wind-wind collision region to the observer.

The bottom panel of Fig~\ref{nt} shows the spectral energy distribution of the predicted anisotropic IC emission for the four different models listed in Table~\ref{models}.
Decreasing $\zeta_{\rm B}$ reduces the maximum momentum of the particles, but increases the normalisation of the non-thermal particles at lower energies. 
Increasing the mass-loss rate of the primary star by a higher factor than those of the secondary star leads to a colliding wind region which is closer to the secondary.
This leads to an increased IC cooling and thus a higher total normalisation of the emission.

The largest flux of the IC emission at $\phi=0.028$ between 20 keV and 30\,keV is reached by model 4 with a flux of $1.81\,10^{-14}\,\mathrm{erg\,s^{-1}\,cm^{-2}}$.
It is still about six times lower than the upper limit on the flux obtained with {\it INTEGRAL} ($F_\mathrm{INTEGRAL}$) and about 2.5 times lower than the upper limit on the flux obtained with {\it NuSTAR} ($F_\mathrm{NuSTAR}$) for $\Gamma=1.5$.

Compared to other works in the literature, the predicted emission at 10\,keV from model 1 is about 3 orders of magnitude lower than that for WR\,140 \citep{PittDough,Reimer}. 
This large difference is due to the powerful WR wind in the WR\,140 system.

Finally, we estimated the kinetic luminosity available in the shock as $L=0.5\,\dot{M}\,v_\infty^2$.
For model~1, the kinetic luminosity of the primary wind is $L_1 =1.27\,10^{36}\,\mathrm{erg\,s^{-1}}$ while that of the secondary wind is $L_2 =1.27\,10^{35}\,\mathrm{erg\,s^{-1}}$.
Considering the wind momentum ratio of this model ($\eta=0.1$), \citet{pittard18} determined the half-opening angles $\theta$ of the shock region (as measured from the secondary star) of the primary and secondary shocks to be 73$^\circ$ and 30$^\circ$, respectively. 
The fraction of wind that is shocked is $f_1 = 0.5 (1 - \cos\theta)=0.35$ for the primary and $f_2 = 0.5 (1+ \cos\theta)=0.93$ for the secondary.
The total kinetic luminosity is thus $L \sim f_1\,L_1 + f_2\,L_2 = 5.6\,10^{34}\,\mathrm{erg\,s^{-1}}$.
The fraction of the wind luminosity that is thermalised is somewhat lower than $f_1$ and $f_2$ since the majority of the pre-shock wind is not normal to the shock.
Considering the above computed fractions thus leads to an upper limit of the IC to kinetic luminosity ratio.
The total kinetic luminosity has to be compared to the full IC emission from model~1 at phase 0.028: $1.2\,10^{31}\,\mathrm{erg\,s^{-1}}$. 
The IC emission is thus about three orders of magnitude lower than the available power.

The radiated luminosity from thermal particles is $L \sim (f_1\,L_1/\chi_1) + (f_2\,L_2/\chi_2)$, where $\chi = v_8^4\,d_{12}/\dot{M}_7$ the cooling parameter for each wind with $v_8$ the terminal velocity in $10^8\,\mathrm{cm\,s^{-1}}$, the distance separating the stars in $10^{12}\,$cm, and $\dot{M}_7$ the mass-loss rate in $10^{-7}\,$M$_{\odot}$\,yr$^{-1}$. 
This leads to $\chi_1=13$ for the primary and $\chi_2=130$ for the secondary of model~1.
The primary wind thus dominates the total luminosity, which is roughly $L \sim 3.4\,10^{34}\,\mathrm{erg\,s^{-1}}$, leading to a flux of $f = 1.25\,10^{-10}\,\mathrm{erg\,s^{-1}\,cm^{-2}}$. 
The unabsorbed flux (i.e. corrected from the ISM hydrogen column density) computed from the {\it XMM-Newton} spectrum at phase 0.019 in the 0.5--10\,keV energy range is about $5\,10^{-11}\,\mathrm{erg\,s^{-1}\,cm^{-2}}$.
The theoretical and observational fluxes are thus in good agreement because the factors $f_1$ and $f_2$ overestimate the wind luminosity that is thermalised.

\section{Summary and conclusions}
\label{summary}
We used X-ray data taken with {\it Swift} and {\it XMM-Newton} to improve the phase coverage of the 2 -- 10\,keV light curve of \stars{}. 
The improved fitting of the X-ray spectra with a 3T plasma model allowed us to achieve a better description of the characteristics of the thermal X-ray emission along the orbit.
Our results indicate a light curve with an almost constant level between phases 0.2 and 0.6 followed by an increase until a maximum around phase 0.8 and then a sharp decay with a minimum around phase 0.1.
This light curve shape is similar to that of other eccentric systems with comparable orbital periods. 
However, unlike the case of HD~166\,734, the X-ray emission of \stars{} preserves a luminosity well above the level of the intrinsic emission of O-type stars, even at minimum. 
This indicates that the colliding wind region in \stars{} is not disrupted during the periastron passage.

We also analysed 3.1\,Ms of {\it INTEGRAL} observations of the {\it EGRET} source 3EG\,2033+4118 corresponding to the Cygnus~OB2 region.
This allowed us to refine the 3$\sigma$ upper limit on the hard X-ray count rate (between 20 and 200\,keV) produced by the sources located within less than $6\arcmin$ from 3EG\,2033+4118.
We then converted this upper limit on the count rate to an upper limit on the flux of the non-thermal emission by modelling a power-law with different spectral indices $\Gamma$.
For $\Gamma=1.5$ (expected for the IC emission), the 3$\sigma$ upper limit on the 20--30\,keV emission from the Cygnus~OB2 region is $1.1\,10^{-13}\,\mathrm{erg\,s^{-1}\,cm^{-2}}$.

Two {\it NuSTAR} observations of \stars{} taken shortly after periastron (phases 0.028 and 0.085) were then investigated to directly search for evidence of an IC hard X-ray emission from the binary thanks to its better angular resolution. 
Combining these spectra with {\it XMM-Newton} data at neighbouring orbital phases, we found no compelling evidence of strong excess above the thermal emission out to energies of 30\,keV. 
Meanwhile, we used the {\it NuSTAR} spectra to infer more stringent upper limits on the flux of the putative non-thermal component.
For $\Gamma=1.5$, the 3$\sigma$ upper limit on the 20--30\,keV emission from \stars{} is $4.6\,10^{-14}\,\mathrm{erg\,s^{-1}\,cm^{-2}}$.

We used the new theoretical model of \citet{pittard20} to predict the anisotropic IC emission from the wind shock region.
The predicted non-thermal fluxes are lower than our observational upper limits, indicating that another quantum leap in sensitivity is required before the non-thermal X-ray emission of O-type star binaries can possibly be detected.

\begin{acknowledgements}
The OHP observing campaigns were supported by the Direction G\'en\'erale de l'Enseignement Non-Obligatoire et de la Recherche Scientifique of the F\'ed\'eration Wallonie-Bruxelles. 
The Li\`ege team thanks the European Space Agency (ESA) and the Belgian Federal Science Policy Office (BELSPO) for their support in the framework of the PRODEX Programme (contracts XMaS and HERMES). 
This work was supported by the Fonds National de la Recherche Scientifique - FNRS, notably under grant n$^\mathrm{o}$ T.0192.19.
This research has made use of the software provided by the High Energy Astrophysics Science Archive Research Center (HEASARC), which is a service of the Astrophysics Science Division at NASA/GSFC and the High Energy Astrophysics Division of the Smithsonian Astrophysical Observatory.
This research has also made use of data from the NuSTAR mission, a project led by the California Institute of Technology, managed by the Jet Propulsion Laboratory, and funded by the National Aeronautics and Space Administration and of the NuSTAR Data Analysis Software (NuSTARDAS) jointly developed by the ASI Science Data Center (ASDC, Italy) and the California Institute of Technology (USA).
We thank the PIs who obtained the Swift, INTEGRAL and XMM-Newton observations used in this work.
\end{acknowledgements}

\begin{appendix}

\section{log of the X-ray observations of \stars{}}

\begin{table}
\centering
  \caption{X-ray observations of \stars{}\label{table_obs}}
\resizebox{0.48\textwidth}{!}{
  \begin{tabular}{cccccc}
\hline
Facility & Date (mid-obs) & Count rate & Exposure & Observation ID/Rev & Phase\,\tablefootmark{a}\\
 & (MJD) & ($\mathrm{count\,s^{-1}}$) & (ks) & & \\
\hline 
\hline
XMM-Newton\,\tablefootmark{b} & 3308.080 & $0.514\pm0.005$& 20.909 & 0200450201/0896 & 0.586\\
 & 3318.059 &$0.392\pm0.004$ & 22.982 & 0200450301/0901 & 0.042\\
 & 3328.043 & $0.485\pm0.004$& 25.056 & 0200450401/0906 & 0.497\\
 & 3338.005 &$0.441\pm0.004$ & 22.982 & 0200450501/0911 & 0.952\\
 & 4219.854 & $0.424\pm0.004$& 31.795 & 0505110301/1353 & 0.207\\
 & 4223.669 &$0.475\pm0.004$ & 33.005 & 0505110401/1355 & 0.381\\
 & 5737.756 &$0.470\pm0.004$ & 29.722 & 0677980601/2114 & 0.497\\
 & 6757.709 & $0.191\pm0.004$& 10.368 & 0740300101/2625 & 0.056\\
 & 7682.729 &$0.179\pm0.004$ & 13.119 & 0780040101/3089 & 0.282\\
 & 7698.880 & $0.093\pm0.002$& 34.668 & 0793183001/3097 & 0.019\\
 & 7856.333 & $0.236\pm0.005$& 8.727 & 0800150101/3176 & 0.206\\
 & 8092.351 & $0.063\pm0.002$ & 22.640 & 0801910601/3294 & 0.956\\
\hline
Swift & 4084.709 & $0.161\pm0.006 $ & 4.486 & 36257003 & 0.038 \\
& 5571.120 & $0.279\pm0.009 $ & 3.170 & 31904001 & 0.890 \\
& 5655.338 & $0.245\pm0.009 $ & 3.349 & 31904002 & 0.734 \\
& 5699.583 & $0.271\pm0.009 $ & 3.592 & 31904003 & 0.754 \\
& 5743.339 & $0.245\pm0.008 $ & 4.159 & 31904004 & 0.752 \\
& 5841.670 & $0.205\pm0.008 $ & 3.469 & 31904005 & 0.240 \\
& 6379.940 & $0.278\pm0.012 $ & 1.832 & 32767001 & 0.811 \\
& 6380.377 & $0.278\pm0.011 $ & 2.504 & 32767002 & 0.831 \\
& 7191.028 & $0.286\pm0.029 $ & 0.350 & 33818001 & 0.836 \\
& 7191.549 & $0.165\pm0.048 $ & 0.073 & 33818002 & 0.860 \\
& 7192.237 & $0.258\pm0.023 $ & 0.504 & 33818003 & 0.891 \\
& 7192.816 & $0.217\pm0.030 $ & 0.244 & 33818005 & 0.918 \\
& 7193.012 & $0.312\pm0.061 $ & 0.083 & 33818004 & 0.927 \\
& 7193.760 & $0.206\pm0.018 $ & 0.662 & 33818006 & 0.961 \\
& 7194.031 & $0.185\pm0.018 $ & 0.568 & 33818007 & 0.973 \\
& 7194.818 & $0.171\pm0.015 $ & 0.736 & 33818008 & 0.009 \\
& 7195.283 & $0.167\pm0.012 $ & 1.131 & 33818009 & 0.030 \\
& 7195.659 & $0.157\pm0.015 $ & 0.707 & 33818010 & 0.048 \\
& 7196.160 & $0.186\pm0.017 $ & 0.672 & 33818011 & 0.070 \\
& 7196.613 & $0.144\pm0.019 $ & 0.415 & 33818012 & 0.091 \\
& 7197.416 & $0.157\pm0.013 $ & 0.878 & 33818013 & 0.128 \\
& 7197.937 & $0.175\pm0.025 $ & 0.269 & 33818014 & 0.152 \\
& 7198.551 & $0.177\pm0.019 $ & 0.492 & 33818016 & 0.180 \\
& 7199.011 & $0.173\pm0.025 $ & 0.284 & 33818017 & 0.201 \\
& 7199.798 & $0.219\pm0.020 $ & 0.535 & 33818018 & 0.237 \\
& 7200.732 & $0.199\pm0.016 $ & 0.814 & 33818020 & 0.279 \\
& 7201.080 & $0.217\pm0.018 $ & 0.669 & 33818021 & 0.295 \\
& 7201.529 & $0.198\pm0.015 $ & 0.842 & 33818022 & 0.316 \\
& 7202.034 & $0.207\pm0.023 $ & 0.401 & 33818023 & 0.339 \\
& 7203.010 & $0.246\pm0.023 $ & 0.455 & 33818024 & 0.383 \\
& 7203.657 & $0.199\pm0.022 $ & 0.402 & 33818025 & 0.413 \\
& 7204.390 & $0.215\pm0.017 $ & 0.754 & 33818026 & 0.446 \\
& 7204.590 & $0.219\pm0.016 $ & 0.863 & 33818027 & 0.455 \\
& 7205.255 & $0.208\pm0.017 $ & 0.716 & 33818028 & 0.486 \\
& 7205.600 & $0.237\pm0.019 $ & 0.683 & 33818029 & 0.501 \\
& 7206.530 & $0.222\pm0.017 $ & 0.751 & 33818031 & 0.544 \\
& 7207.251 & $0.218\pm0.018 $ & 0.660 & 33818032 & 0.577 \\
& 7207.596 & $0.237\pm0.017 $ & 0.780 & 33818033 & 0.592 \\
& 7208.861 & $0.208\pm0.015 $ & 0.879 & 33818035 & 0.650 \\
& 7209.458 & $0.212\pm0.016 $ & 0.813 & 33818036 & 0.677 \\
& 7210.750 & $0.218\pm0.021 $ & 0.482 & 33818038 & 0.736 \\
& 7211.454 & $0.285\pm0.021 $ & 0.663 & 33818039 & 0.769 \\
& 7211.988 & $0.261\pm0.018 $ & 0.775 & 33818040 & 0.793 \\
& 7212.375 & $0.238\pm0.017 $ & 0.806 & 33818041 & 0.811 \\
& 7212.708 & $0.267\pm0.018 $ & 0.786 & 33818042 & 0.826 \\
& 7213.790 & $0.251\pm0.018 $ & 0.749 & 33818044 & 0.875 \\
& 7213.852 & $0.222\pm0.020 $ & 0.573 & 33818043 & 0.878 \\
& 7284.444 & $0.137\pm0.007 $ & 2.772 & 32767003 & 0.100 \\
& 7510.027 & $0.233\pm0.008 $ & 3.494 & 34282006\,\tablefootmark{c} & 0.398 \\
& 7524.601 & $0.176\pm0.007 $ & 3.329 & 34282008\,\tablefootmark{c} & 0.063 \\
& 7540.490 & $0.258\pm0.011 $ & 2.073 & 34282010\,\tablefootmark{c} & 0.789 \\
& 7591.626 & $0.146\pm0.005 $ & 4.922 & 34282078\,\tablefootmark{c} & 0.123 \\
& 7618.558 & $0.208\pm0.018 $ & 0.662 & 34282022 & 0.352 \\
& 7879.050 & $0.187\pm0.016 $ & 0.737 & 93146003 & 0.243 \\
& 7907.287 & $0.226\pm0.011 $ & 1.730 & 93146005 & 0.532 \\
& 7921.106 & $0.175\pm0.008 $ & 2.934 & 93146006\,\tablefootmark{c} & 0.163 \\
& 7935.656 & $0.283\pm0.011 $ & 2.170 & 93146007\,\tablefootmark{c} & 0.827 \\
& 7942.261 & $0.211\pm0.007 $ & 3.846 & 34282018\,\tablefootmark{c} & 0.129 \\
& 7948.354 & $0.182\pm0.010 $ & 1.973 & 34282081\,\tablefootmark{c} & 0.407 \\
& 7951.202 & $0.191\pm0.010 $ & 1.898 & 34282084\,\tablefootmark{c} & 0.537 \\
& 8095.676 & $0.156\pm0.008 $ & 2.664 & 10451001 & 0.132 \\
& 8105.269 & $0.202\pm0.008 $ & 2.878 & 10451002 & 0.570 \\
& 8115.631 & $0.150\pm0.008 $ & 2.619 & 10451003 & 0.043 \\
& 8125.724 & $0.192\pm0.008 $ & 2.741 & 10451004 & 0.503 \\
& 8135.195 & $0.219\pm0.009 $ & 2.721 & 10451005 & 0.936 \\
& 8145.287 & $0.174\pm0.008 $ & 2.691 & 10451006 & 0.397 \\
& 8155.291 & $0.262\pm0.010 $ & 2.443 & 10451007 & 0.853 \\
& 8166.120 & $0.198\pm0.009 $ & 2.479 & 10451008 & 0.347 \\
& 8175.705 & $0.228\pm0.009 $ & 2.882 & 10451009 & 0.785 \\
& 8185.109 & $0.202\pm0.010 $ & 1.993 & 10451010 & 0.214 \\
& 8229.108 & $0.206\pm0.022 $ & 0.432 & 34282137 & 0.223 \\
& 8355.695 & $0.151\pm0.010 $ & 1.533 & 88806001 & 0.001 \\
& 8357.961 & $0.172\pm0.011 $ & 1.380 & 88807001 & 0.105 \\
\hline
NuSTAR\,\tablefootmark{d} & 8356.290 & $0.047\pm0.031$& 40.919 & 30410001002 & 0.028\\
 & 8357.527 & $0.074\pm0.051$& 50.030 & 30410002002 & 0.085\\
\hline
\end{tabular}
  }
  \tablefoot{
\tablefoottext{a} {Phases were computed with the orbital solution revised in Sect.~\ref{orb_sol}.}
\tablefoottext{b} {The count rates are in the 2--10\,keV energy band for EPIC/pn except for MJD~8092.351 because \stars{} was observed with a large off-axis angle and EPIC/pn was in timing mode making only MOS2 data exploitable.}
\tablefoottext{c} {The count rates of these observations were taken from the Swift light curve repository.}
\tablefoottext{d} {The count rates are for FPMA.}
}
  \end{table}
  
\section{Results of the spectral fitting of the X-ray spectra.}
\begin{table*}
  \caption{Results of the X-ray spectral fittings with the dustscat$\times$TBnew$\times$phabs$\times$(apec+apec+apec) model. \label{table_fit}}
\centering
\resizebox{0.85\textwidth}{!}{
\begin{tabular}{cccccccccc}
\hline
Instrument & Phase & $\mathrm{N_H^{add}}$ & $kT_1$ & $n_1$ & $kT_2$ & $n_2$ & $kT_3$ & $n_3$ & $\chi^2$ (dof)\\
 & & ($10^{22}\,\mathrm{cm^{-2}}$) & (keV) & ($10^{-2}\,\mathrm{cm^{-5}}$) & (keV) & ($10^{-3}\,\mathrm{cm^{-5}}$) & (keV) & ($10^{-3}\,\mathrm{cm^{-5}}$) & \\
\hline 
\hline
{\it XMM-Newton} & $0.586$ & $0.65^{+0.06}_{-0.05}$ & $0.37^{+0.06}_{-0.05}$ & $1.46^{+0.9}_{-0.4}$ & $0.86^{+0.04}_{-0.03}$ & $8.8\pm1.4$ & $2.10^{+0.08}_{-0.07}$ & $7.0\pm0.4$ &  1.12 (785)\\
 & $0.042$ & $0.89^{+0.08}_{-0.07}$ & $0.30^{+0.10}_{-0.02}$ & $2.10^{+0.9}_{-1.3}$ & $0.77\pm0.02$ & $12.3\pm1.0$ & $1.95^{+0.10}_{-0.08}$ & $5.1\pm0.5$ &  1.13 (709)\\
 & $0.497$ & $0.61^{+0.07}_{-0.03}$ & $0.48^{+0.45}_{-0.10}$ & $1.10\pm0.1$ & $0.88^{+0.04}_{-0.10}$ & $6.9^{+3.2}_{-0.7}$ & $2.21^{+0.09}_{-0.14}$ & $6.1^{+0.7}_{-0.4}$ &  1.17 (837)\\
 & $0.952$ & $0.83^{+0.05}_{-0.04}$ & $0.28^{+0.20}_{-0.04}$ & $3.57^{+2.2}_{-0.9}$ & $0.78^{+0.03}_{-0.02}$ & $12.6^\pm1.0$ & $1.98^{+0.10}_{-0.09}$ & $5.7\pm0.5$ &  1.34 (731)\\
 & $0.207$ & $0.79\pm0.04$ & $0.31^{+0.08}_{-0.02}$ & $2.33^{+0.8}_{-1.1}$ & $0.80\pm0.02$ & $11.1\pm1.0$ & $2.09^{+0.10}_{-0.09}$ & $5.2\pm0.4$ &  1.28 (889)\\
 & $0.381$ & $0.67^{+0.05}_{-0.04}$ & $0.33^{+0.05}_{-0.06}$ & $1.80^{+0.8}_{-0.4}$ & $0.81^{+0.03}_{-0.02}$ & $9.9\pm0.1$ & $2.03^{+0.07}_{-0.06}$ & $6.8\pm0.4$ &  1.19 (1018)\\
 & $0.497$ & $0.58^{+0.07}_{-0.04}$ & $0.50^{+0.12}_{-0.14}$ & $1.08\pm0.2$ & $0.82^{+0.30}_{-0.10}$ & $6.2^{+3.5}_{-0.9}$ & $2.10^{+0.28}_{-0.13}$ & $6.6^{+0.9}_{-1.7}$ &  1.16 (436)\\
 & $0.056$ & $0.96^{+0.08}_{-0.10}$ & $0.29^{+0.27}_{-0.06}$ & $2.51^{+3.2}_{-1.6}$ & $0.74^{+0.40}_{-0.30}$ & $1.4\pm0.2$ & $2.34^{+0.60}_{-0.30}$ & $3.6\pm0.8$ &  1.18 (261)\\
 & $0.282$ & $0.90^{+0.07}_{-0.08}$ & $0.29^{+0.05}_{-0.04}$ & $4.36^{+3.0}_{-2.0}$ & $0.91^{+0.05}_{-0.08}$ & $12.6\pm2.0$ & $2.48^{+0.90}_{-0.40}$ & $3.7^{+1.1}_{-1.3}$ &  1.58 (324)\\
 & $0.019$ & $0.92^{+0.09}_{-0.10}$ & $0.33^{+0.13}_{-0.05}$ & $2.30^{+2.2}_{-1.3}$ & $0.85^{+0.07}_{-0.05}$ & $10.9\pm2.0$ & $2.17^{+0.30}_{-0.20}$ & $5.4^{+0.9}_{-0.8}$ &  0.96 (175)\\
 & $0.206$ & $0.82\pm0.10$ & $0.34^{+0.24}_{-0.08}$ & $1.88^{+2.1}_{-0.8}$ & $0.78^{+0.09}_{-0.04}$ & $12.5^{+3.0}_{-6.0}$ & $1.95^{+0.19}_{-0.15}$ & $5.8\pm0.9$ &  1.30 (283)\\
 & $0.956$ & $0.83^{+0.20}_{-0.15}$ & $0.26^{+0.09}_{-0.19}$ & $4.81^{+2.37}_{-8.66}$ & $0.91\pm0.11$ & $11.7^{+3.3}_{-4.6}$ & $2.10^{+0.31}_{-1.14}$ & $5.91^{+3.06}_{-2.07}$ &  1.46 (81)\\
 \hline
{\it XMM-Newton} & $0.586$ & $0.69\pm0.03$ & [0.34] & $1.8\pm0.3$ & [0.83] & $10.2\pm0.5$ & [2.13] & $6.5\pm0.2$ &  1.13 (788)\\
 & $0.042$ & $0.91\pm0.03$ & \rule[0.5ex]{2.5em}{0.55pt} & $2.0\pm0.3$ & \rule[0.5ex]{2.5em}{0.55pt} & $11.9\pm0.5$ & \rule[0.5ex]{2.5em}{0.55pt} & $4.0\pm0.2$ &  1.18 (712)\\
 & $0.497$ & $0.71\pm0.01$ & \rule[0.5ex]{2.5em}{0.55pt} & $2.0^{+0.3}_{-0.2}$ & \rule[0.5ex]{2.5em}{0.55pt} & $9.3\pm0.5$ & \rule[0.5ex]{2.5em}{0.55pt} & $6.1\pm0.2$ &  1.19 (840)\\
 & $0.952$ & $0.81\pm0.02$ & \rule[0.5ex]{2.5em}{0.55pt} & $2.4\pm0.3$ & \rule[0.5ex]{2.5em}{0.55pt} & $11.8\pm0.6$ & \rule[0.5ex]{2.5em}{0.55pt} & $4.8\pm0.2$ &  1.39 (734)\\
 & $0.207$ & $0.78\pm0.02$ & \rule[0.5ex]{2.5em}{0.55pt} & $2.1\pm0.2$ & \rule[0.5ex]{2.5em}{0.55pt} & $10.6\pm0.5$ & \rule[0.5ex]{2.5em}{0.55pt} & $4.9\pm0.2$ &  1.29 (892)\\
 & $0.381$ & $0.70\pm0.02$ & \rule[0.5ex]{2.5em}{0.55pt} & $1.9\pm0.2$ & \rule[0.5ex]{2.5em}{0.55pt} & $10.6\pm0.4$  & \rule[0.5ex]{2.5em}{0.55pt}& $5.9\pm0.1$ &  1.21 (1021)\\
 & $0.497$ & $0.71\pm0.03$ & \rule[0.5ex]{2.5em}{0.55pt} & $2.1\pm0.3$ & \rule[0.5ex]{2.5em}{0.55pt} & $9.7\pm0.6$ & \rule[0.5ex]{2.5em}{0.55pt} & $6.2\pm0.2$ &  1.22 (439)\\
 & $0.056$ & $0.96\pm0.05$ & \rule[0.5ex]{2.5em}{0.55pt} & $2.5\pm0.5$ & \rule[0.5ex]{2.5em}{0.55pt} & $11.2\pm1.1$ & \rule[0.5ex]{2.5em}{0.55pt} & $3.6\pm0.4$ &  1.22 (264)\\
 & $0.282$ & $0.83\pm0.05$ & \rule[0.5ex]{2.5em}{0.55pt} & $2.3\pm0.5$ & \rule[0.5ex]{2.5em}{0.55pt} & $5.5\pm1.0$ & \rule[0.5ex]{2.5em}{0.55pt} & $4.7\pm0.3$ &  1.57 (327)\\
 & $0.019$ & $0.92\pm0.06$ & \rule[0.5ex]{2.5em}{0.55pt} & $2.1\pm0.6$ & \rule[0.5ex]{2.5em}{0.55pt} & $11.3\pm1.2$ & \rule[0.5ex]{2.5em}{0.55pt} & $5.3\pm0.4$ &  0.95 (176)\\
 & $0.206$ & $0.86^{+0.04}_{-0.05}$ & \rule[0.5ex]{2.5em}{0.55pt} & $2.4\pm0.5$ & \rule[0.5ex]{2.5em}{0.55pt} & $12.6\pm1.0$ & \rule[0.5ex]{2.5em}{0.55pt} & $4.6\pm0.3$ &  1.31 (285)\\
 & $0.956$ & $0.77\pm0.09$ & \rule[0.5ex]{2.5em}{0.55pt} & $1.9^{+0.8}_{-0.7}$ & \rule[0.5ex]{2.5em}{0.55pt} & $11.5\pm2.2$ & \rule[0.5ex]{2.5em}{0.55pt} & $6.2\pm0.7$ &  1.46 (84)\\
 \hline 
{\it XMM-Newton} & $0.586$ & $0.71\pm0.02$ & [0.34]  & [2.11] & [0.83] &  $9.4\pm0.5$ & [2.13] & $6.9\pm0.2$ & 1.13 (789)\\
 & $0.042$ & $0.91\pm0.02$ & \rule[0.5ex]{2.5em}{0.55pt} & \rule[0.5ex]{2.5em}{0.55pt} & \rule[0.5ex]{2.5em}{0.55pt}  & $11.5\pm0.5$ & \rule[0.5ex]{2.5em}{0.55pt} & $4.3\pm0.2$ &1.16 (713)\\
 & $0.497$ & $0.71\pm0.01$ & \rule[0.5ex]{2.5em}{0.55pt} & \rule[0.5ex]{2.5em}{0.55pt} & \rule[0.5ex]{2.5em}{0.55pt}  & $8.8\pm0.4$ & \rule[0.5ex]{2.5em}{0.55pt} & $6.5\pm0.2$ & 1.18 (841)\\
 & $0.952$ & $0.79\pm0.02$ & \rule[0.5ex]{2.5em}{0.55pt} & \rule[0.5ex]{2.5em}{0.55pt} & \rule[0.5ex]{2.5em}{0.55pt}  & $11.9\pm0.5$ & \rule[0.5ex]{2.5em}{0.55pt} & $5.0\pm0.2$ & 1.37 (735)\\
 & $0.207$ & $0.78\pm0.01$ & \rule[0.5ex]{2.5em}{0.55pt} & \rule[0.5ex]{2.5em}{0.55pt} & \rule[0.5ex]{2.5em}{0.55pt}  & $10.3\pm0.4$ & \rule[0.5ex]{2.5em}{0.55pt} & $5.2\pm0.2$ & 1.28 (893)\\
 & $0.381$ & $0.71\pm0.01$ & \rule[0.5ex]{2.5em}{0.55pt} & \rule[0.5ex]{2.5em}{0.55pt} & \rule[0.5ex]{2.5em}{0.55pt}  & $10.1\pm0.4$ & \rule[0.5ex]{2.5em}{0.55pt} & $6.3^{+0.1}_{-0.2}$ & 1.20 (1022)\\
 & $0.497$ & $0.70\pm0.02$ & \rule[0.5ex]{2.5em}{0.55pt} & \rule[0.5ex]{2.5em}{0.55pt} & \rule[0.5ex]{2.5em}{0.55pt}  & $9.2\pm0.5$ & \rule[0.5ex]{2.5em}{0.55pt} & $6.6\pm0.2$ & 1.19 (440)\\
 & $0.056$ & $0.93\pm0.03$ & \rule[0.5ex]{2.5em}{0.55pt} & \rule[0.5ex]{2.5em}{0.55pt} & \rule[0.5ex]{2.5em}{0.55pt}  & $11.6\pm0.9$ & \rule[0.5ex]{2.5em}{0.55pt} & $3.7\pm0.4$ & 1.22 (265)\\
 & $0.282$ & $0.81\pm0.02$ & \rule[0.5ex]{2.5em}{0.55pt} & \rule[0.5ex]{2.5em}{0.55pt} & \rule[0.5ex]{2.5em}{0.55pt}  & $11.9\pm0.8$ & \rule[0.5ex]{2.5em}{0.55pt} & $5.0\pm0.3$ & 1.57 (328)\\
 & $0.019$ & $0.91\pm0.03$ & \rule[0.5ex]{2.5em}{0.55pt} & \rule[0.5ex]{2.5em}{0.55pt} & \rule[0.5ex]{2.5em}{0.55pt}  & $11.1\pm1.0$ & \rule[0.5ex]{2.5em}{0.55pt} & $5.6\pm0.4$ & 0.94 (179)\\
 & $0.206$ & $0.84\pm0.03$ & \rule[0.5ex]{2.5em}{0.55pt} & \rule[0.5ex]{2.5em}{0.55pt} & \rule[0.5ex]{2.5em}{0.55pt}  & $12.6\pm0.9$ & \rule[0.5ex]{2.5em}{0.55pt} & $4.9\pm0.3$ & 1.30 (287)\\
 & $0.956$ & $0.78\pm0.06$ & \rule[0.5ex]{2.5em}{0.55pt} & \rule[0.5ex]{2.5em}{0.55pt} & \rule[0.5ex]{2.5em}{0.55pt}  & $11.0^{+1.9}_{-1.8}$ & \rule[0.5ex]{2.5em}{0.55pt} & $6.5\pm0.7$ & 1.45 (85)\\
\hline
{\it Swift} & $0.95-0.05$ & $1.19\pm0.16$ & [0.34]  & [2.11] & [0.83] & $9.6^{+3.2}_{-3.1}$& [2.13] & $5.5\pm0.9$ &1.29 (83)\\
& $0.05-0.15$ & $0.96\pm0.16$ & \rule[0.5ex]{2.5em}{0.55pt} &  \rule[0.5ex]{2.5em}{0.55pt} &  \rule[0.5ex]{2.5em}{0.55pt} & $7.8^{+3.6}_{-3.4} $ & \rule[0.5ex]{2.5em}{0.55pt} & $5.1^{+1.2}_{-1.3} $ & 1.92 (68)\\
& $0.15-0.25$ & $1.04^{+0.17}_{-0.16} $ & \rule[0.5ex]{2.5em}{0.55pt} &  \rule[0.5ex]{2.5em}{0.55pt} &  \rule[0.5ex]{2.5em}{0.55pt} & $12.1^{+4.4}_{-4} $ & \rule[0.5ex]{2.5em}{0.55pt} & $6.5^{+1.3}_{-1.4} $ & 1.08 (62)\\
& $0.25-0.35$ & $1.14^{+0.22}_{-0.21} $ & \rule[0.5ex]{2.5em}{0.55pt} &  \rule[0.5ex]{2.5em}{0.55pt} &  \rule[0.5ex]{2.5em}{0.55pt} & $15.5^{+6}_{-5.7} $ & \rule[0.5ex]{2.5em}{0.55pt} & $5.4\pm2.0$ &1.30 (40)\\
& $0.35-0.45$ & $1.19^{+0.22}_{-0.23} $ & \rule[0.5ex]{2.5em}{0.55pt} &  \rule[0.5ex]{2.5em}{0.55pt} &  \rule[0.5ex]{2.5em}{0.55pt} & $13.8^{+6.1}_{-5.8} $ & \rule[0.5ex]{2.5em}{0.55pt} & $6.2\pm2.0$ &1.45 (44)\\
& $0.45-0.55$ & $0.89^{+0.17}_{-0.19} $ & \rule[0.5ex]{2.5em}{0.55pt} &  \rule[0.5ex]{2.5em}{0.55pt} &  \rule[0.5ex]{2.5em}{0.55pt} & $9.4^{+4.6}_{-4.5} $ & \rule[0.5ex]{2.5em}{0.55pt} & $7.9\pm1.5$ &0.61 (64)\\
& $0.55-0.65$ & $1.08^{+0.21}_{-0.22} $ & \rule[0.5ex]{2.5em}{0.55pt} &  \rule[0.5ex]{2.5em}{0.55pt} &  \rule[0.5ex]{2.5em}{0.55pt} & $12.8^{+5.6}_{-5.5} $ & \rule[0.5ex]{2.5em}{0.55pt} & $6.8\pm1.8$ &1.05 (36)\\
& $0.65-0.75$ & $1.14^{+0.18}_{-0.19} $ & \rule[0.5ex]{2.5em}{0.55pt} &  \rule[0.5ex]{2.5em}{0.55pt} &  \rule[0.5ex]{2.5em}{0.55pt} & $17.7^{+6.1}_{-5.9} $ & \rule[0.5ex]{2.5em}{0.55pt} & $6.9\pm2.0$ &1.00 (51)\\
& $0.75-0.85$ & $0.85^{+0.09}_{-0.10} $ & \rule[0.5ex]{2.5em}{0.55pt} &  \rule[0.5ex]{2.5em}{0.55pt} &  \rule[0.5ex]{2.5em}{0.55pt} & $12.8^{+3.1}_{-3} $ & \rule[0.5ex]{2.5em}{0.55pt} & $8.7\pm1.1$ &1.08 (163)\\
& $0.85-0.95$ & $1.10^{+0.09}_{-0.12} $ & \rule[0.5ex]{2.5em}{0.55pt} &  \rule[0.5ex]{2.5em}{0.55pt} &  \rule[0.5ex]{2.5em}{0.55pt} & $18.1^{+4.1}_{-3.9} $ & \rule[0.5ex]{2.5em}{0.55pt} & $6.7^{+1.2}_{-1.3} $ & 1.44 (105)\\
\hline
{\it Nustar} & $0.028$ & [0.91] & [0.34] & [2.11] & [0.83] & [11.8] & [2.13] & $3.3\pm0.3$ & 0.89 (27)\\
 & $0.085$ & [0.90] & \rule[0.5ex]{2.5em}{0.55pt} & \rule[0.5ex]{2.5em}{0.55pt} & \rule[0.5ex]{2.5em}{0.55pt} & [11.2] & \rule[0.5ex]{2.5em}{0.55pt} & $3.7\pm0.4$  & 0.84 (21)\\
\hline
\end{tabular}
}
\tablefoot{The values in the square brackets are fixed in the models. 
\rule[0.5ex]{2.em}{0.4pt} means that the value is the same as in the previous line.
The errors on the spectral parameters are given by the 90\% confidence level of the marginal distribution of the values taken by the parameters during the MCMC analysis.}
\end{table*}

\section{Modelisation of the wind column density}
\label{williams_model}
\begin{figure}[h]
\begin{center}
  \resizebox{9cm}{!}{\includegraphics{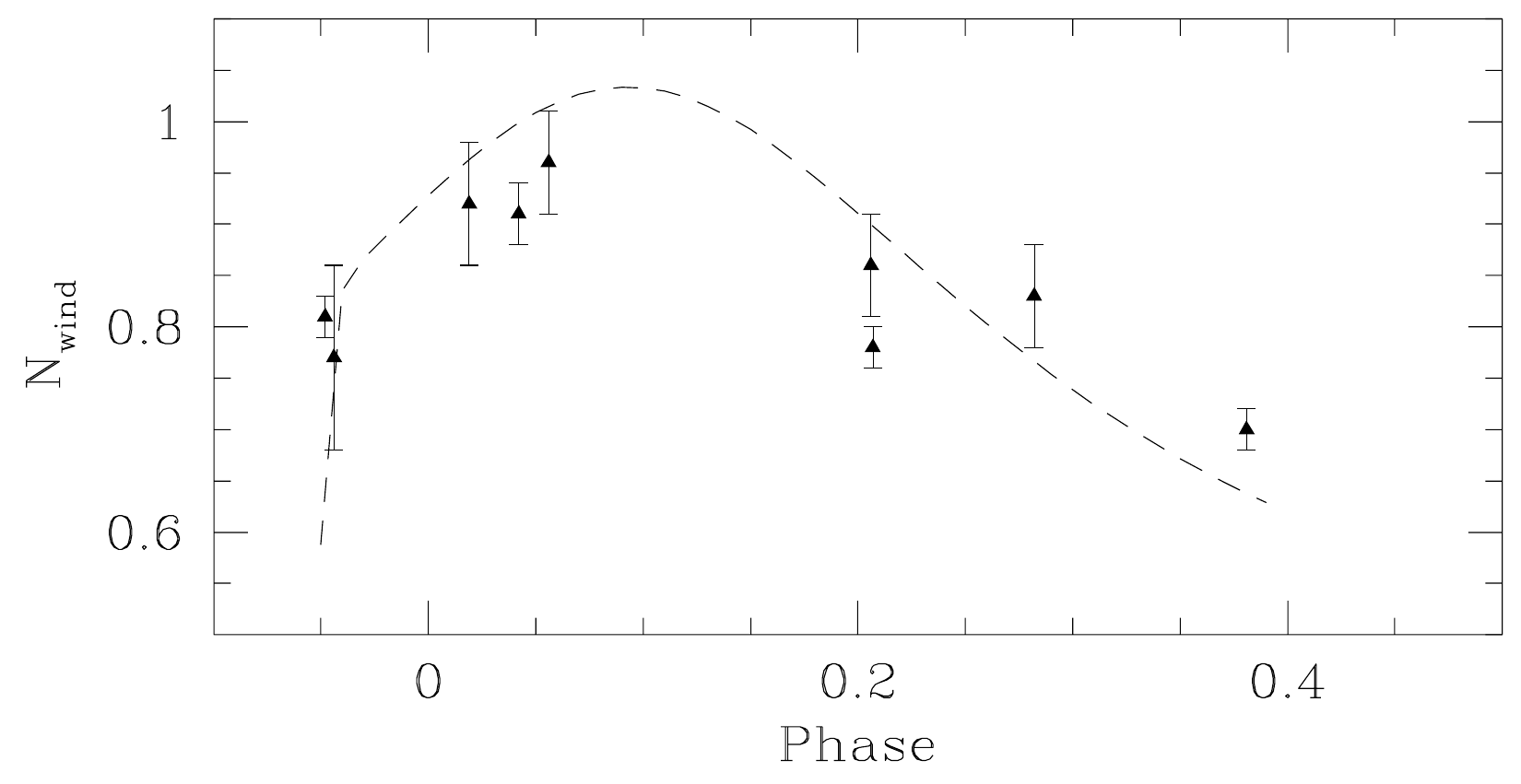}}
\end{center}  
\caption{Variations in the column density in $10^{22}\,\mathrm{cm^{-2}}$ obtained in the fits to the {\it XMM-Newton} spectra (triangular symbols with error bars) fitted with the model of \citet{wil90} for $i = 20^{\circ}$ (dashed line).}
\label{columnwil90}
\end{figure}

We adapted the model of \citet{wil90} to evaluate the line-of-sight column density of the stellar winds as a function of orbital phase. 
For an orbital inclination $i \leq 90^{\circ}$, we have
\begin{equation}
  N_{\rm wind}(\phi) = \frac{N_0}{\frac{r}{a}\,|\cos{(v + \omega)}|\sin{i}\,\theta\,\sqrt{A}}\,\left(\arctan{\frac{u_2}{\theta}} - \arctan{\frac{u_1}{\theta}}\right),
\end{equation}where $N_0$ is a scaling parameter depending on the wind density, $v$ is the true anomaly, $u_1 = \frac{B}{2\,\sqrt{A}}$, $u_2 = \sqrt{A}\,\tan{\psi}+\frac{B}{2\,\sqrt{A}}$, with
\begin{eqnarray}
  A & = & 1 + \cot^2{i}\,\tan^2{(v + \omega)} \\
  B & = & 2\,|\tan{(v + \omega)}|\,\cot^2{i} \\
  C & = & 1 + \cot^2{i} \\
  \theta^2 & = & C - \frac{B^2}{4\,A} 
  \end{eqnarray}
and $\psi$ is given by
\begin{eqnarray}
  \psi  = \frac{\pi}{2} + (v + \omega) & {\rm if} & (v + \omega) \in [0,\frac{\pi}{2}]\\
  \psi = \frac{3\,\pi}{2} - (v + \omega) & {\rm if} & (v + \omega) \in [\frac{\pi}{2},\frac{3\,\pi}{2}]\\
  \psi = (v + \omega) - \frac{3\,\pi}{2} & {\rm if} & (v + \omega) \in [\frac{3\,\pi}{2},2\,\pi]
.\end{eqnarray}
For \stars\ all parameters but $i$ and $N_0$ are known from the orbital solution. The value 
$N_0$ can be determined requiring that the mean value of the computed $N_{\rm wind}$ equals the mean value of the observed values in the relevant phase interval. 
We then computed the $\chi^2$ as a function of orbital inclination by comparing the model to the column density values determined from the {\it XMM-Newton} spectra. 
The lowest $\chi^2$ is reached for $i = \left(20^{+6}_{-4}\right)$\,$^{\circ}$ and the best fit is shown in Fig.\,\ref{columnwil90}. 
Whereas the trend with phase appears well reproduced, this inclination is significantly lower than our estimate based on the typical masses of the stars and would imply unrealistically high masses of 190 and 150\,M$_{\odot}$ for the stars.  

The most likely explanation for this discrepancy is that the model underestimates $i$ because it assumes that the emission arises from a single point at the shock apex. 
In reality, however, the X-ray emitting region is extended and the wind column density inferred from the observations represents the mean of the columns along the various sightlines towards this extended region. 
Another limitation of the model is that this does not account for Coriolis deflections of the shock cone, which might also affect the phase-dependence of $N_{\rm wind}$.

\end{appendix}

\end{document}